\documentclass[final,5p,times,twocolumn]{elsarticle}

\usepackage{url}            
\usepackage[T1]{fontenc}    
\usepackage[utf8]{inputenc} 
\usepackage{xcolor}

\usepackage{graphicx}
\usepackage{amsmath, amssymb, amsfonts}
\usepackage{mathtools}
\usepackage{url, hyperref}

 \usepackage{ulem}

\DeclareMathOperator{\sech}{sech}

\journal{Physica D: Nonlinear Phenomena}

\begin{document}

\begin{frontmatter}

\title{ A bound state attractor in optical turbulence}

\author[InPhyNi]{Cl\'ement Coll\'eaux}
\author[AstonMaths]{Jonathan Skipp}
\author[InPhyNi]{Sergey Nazarenko}
    \ead{sergey.NAZARENKO@univ-cotedazur.fr}
\author[AstonMaths]{Jason Laurie}
\affiliation[InPhyNi]{organization={Universit\'e C\^ote d'Azur, CNRS-Institut de Physique de Nice},
            addressline={17 Rue Julien Laupr\^etre}, 
            city={Nice},
            postcode={06200},
            country={France}}
\affiliation[AstonMaths]{organization={Department of Applied Mathematics and Data Science, College of Engineering and Physical Sciences, Aston University},
            addressline={Aston Triangle}, 
            city={Birmingham},
            postcode={B4 7ET}, 
            country={United Kingdom}}

\begin{abstract}
We study numerically the nonintegrable dynamics of coherent, solitonic, nonlinear waves, in a spatially nonlocal nonlinear Schr\"odinger equation  relevant to realistic modelling of optical systems: the Schr\"odinger-Helmholtz equation.
We observe a single oscillating, coherent solitary wave emerging from a variety of initial conditions. 
Using the direct scattering transform of the (integrable) cubic nonlinear Schr\"odinger equation, we find that this structure is a bound state, comprising of a primary and secondary soliton whose amplitudes oscillate 
in phase opposition.
We interpret this as the solitons periodically exchanging mass. 
We also observe bound states comprising of three oscillating solitons, hinting at the existence of a family of multi-soliton bound states.
Focusing on the two-soliton bound state, 
we observe it self-organising from an initial state of incoherent turbulence, and from solitonic structures launched into the system. 
When a single (primary) solitonic structure is launched, a resonance process between it and waves in the system generates the secondary soliton, resulting in the bound state.
Further, when two solitons are initially launched, we show that they can merge if their phases are synchronised when they collide. 
When the system is launched from a turbulent state comprised of many initial solitons,
we propose that the bound state formation is preceded by a sequence of binary collisions, in which the mass is transferred on average from the weak soliton to the strong one, with occasional soliton mergers. Both processes lead to increasingly stronger and fewer dominant solitons. 
The final state---a solitary bound state surrounded by weakly nonlinear waves---is robust and ubiquitous. We propose that for nonlocal media, 
a bound state comprising of at least two solitons
is a more typical statistical attractor than 
the
single-soliton attractor suggested in previous literature.
\end{abstract}




\end{frontmatter}


\section{Introduction}
\label{sec:intro}

A broad class of optical systems, in which 
quasi-monochromatic light propagates through nonlinear media, 
exhibit solitons:
coherent, solitary, strongly nonlinear waves, which balance wave dispersion with nonlinear self-focusing, and thereby translate through the system with no overall change of shape.
Such systems can be modelled, to the first approximation, by the one-dimensional, focusing, nonlinear Schr{\"o}dinger equation (NLSE),
\begin{equation}
    \label{eq:NLSE}
    i \frac{\partial u}{\partial t} + \frac{1}{2} \frac{\partial^2 u}{\partial x^2} + u|u|^2 = 0.
\end{equation}
Here $u(x,t) \in  \mathbb{C}$ is the envelope of the electric field of the light inside the medium. 
In the case of light passing through an optical sample, the timelike variable $t$ represents the distance along the beam axis, and $x$ is the transverse spatial coordinate (here we consider systems with one spatial dimension). In optical fibres, $t$ represents the longitudinal distance and $x$ is the physical time.
The nonlinear term in Eq.~\eqref{eq:NLSE} arises from the Kerr effect: the spatially local refractive index change due to the intensity $|u|^2$ of the input beam~\citep{newellmoloney1992nonlinearopticsbook, boyd2008nonlinearopticsbook}.

The single-soliton solution of the NLSE in infinite space is 
\begin{equation}
    \label{eq:NLSE_soli}
    u(x,t) = A \sech \left[ A(x-s-vt)\right]
                e^{iv(x-s)}e^{i(A^2-v^2)t/2}e^{i\phi},
\end{equation}
where $A$ is the soliton amplitude, $v$ its velocity, and $s$, $\phi$ its initial position and phase.
The NLSE is integrable, a consequence of which is that solitons collide elastically: when they collide they preserve their shape, speed and amplitude, undergoing only a change of phase~\citep{Zabusky_65}.

However, in real physical systems, perfect integrability is broken due to subleading physical effects that introduce new nonlinear terms to the NLSE. This deviation from integrability leads to a richer variety of soliton dynamics. Solitons may become inelastic, i.e.\ they can become strongly modified or even merge upon collision, and may interact strongly with the background field of weakly nonlinear waves. This is particularly relevant in the context of nonintegrable optical wave turbulence, studied in experimentally and theoretically in Refs.~\citep{Bortolozzo2009optical, Laurie_12}. There, it was observed that an initially turbulent state, consisting of multiple solitons propagating on a background of weakly nonlinear waves, evolves via a sequence of inelastic soliton collisions towards a state in which one single dominant coherent wave survives, having absorbed all the others. The final state is one dominant, solitonic structure, surrounded by small amplitude, weakly nonlinear waves.

The tendency for solitonic structures to coalesce into a single dominant structure coexisting with weakly nonlinear waves in nonintegrable nonlinear Schr{\"o}dinger systems was first described by Zakharov {\it et al.}~\citep{zakharov1988soliton}. They termed this scenario ``soliton turbulence'', although it must be emphasised that in the strictest sense, solitons cannot be defined for nonintegrable systems, see Sec.~\ref{subsec:soli_definition} below. Additionally, they suggested that the single dominant coherent structure surrounded by small-scale waves is a ``statistical attractor'', in the sense of a universal end state that arbitrary initial conditions evolve towards.
Using statistical-mechanics arguments, it was determined in~\citep{Jordan2000meanfield, Rumpf2001coherent} that the final large soliton is a minimiser of the energy in a microcanonical ensemble, with the small-scale waves in the final state acting as a reservoir of excess energy that is present in the initial condition.

In this paper we characterise the statistical attracting state, and examine the processes that lead to it, in a spatially nonlocal variant of the NLSE, revisiting the scenario of optical wave turbulence described in Ref.~\citep{Laurie_12}.
The model we focus on incorporates deviations from the NLSE where the change of refractive index responds nonlocally to the input beam. For example, in thermo-optic crystals heating by the beam spreads through through the crystal by diffusion~\citep{castillo1996formation, Bekenstein2015_OpticalNewtSchro, Faccio2016_OpticalNewtSchro}, or in elasto-optic media such as liquid crystals, the input beam reorients the liquid crystal molecules, and the reorientation spreads by long-range elastic forces~\citep{conti2003route, peccianti2003optical, Bortolozzo2009optical}.
Such systems can be modelled by the Schr{\"o}dinger-Helmholtz equation (SHE),
\begin{subequations}\label{eq:SHE}
\begin{align}
    i \frac{\partial u}{\partial t}
    + \frac{1}{2} \frac{\partial^2 u}{\partial x^2}
    + Vu  &= 0,\label{eq:SHE_schrodinger}\\
    \left( 1 - \beta \frac{\partial^2}{\partial x^2}\right)V &= |u|^2\label{eq:SHE_helmholtz}.
\end{align}
\end{subequations}
This system is so called because the of change refractive index $V\left(|u(x,t)|^2\right)$ solves the Helmholtz equation~\eqref{eq:SHE_helmholtz}, which incorporates both spatially local and nonlocal effects, the latter controlled by the nonlocality parameter $\beta$. Evidently, this parameter also controls the nonintegrability as sending $\beta\to 0$ in the SHE recovers the NLSE. 
The nonintegrability of the SHE was demonstrated in~\citep{Laurie_12}, as it supports resonant nonlinear six-wave interactions.

Examining the dynamics of the SHE, we show that the dominant final coherent structure is in fact comprised of a bound state of 
multiple
spatiotemporally coincident solitons that periodically exchange mass, defined in Eq.~\eqref{eq:waveaction}, leading to a pulsating peak (see Sec.~\ref{subsec:soli_definition} regarding the usage of ``soliton'' in this context). 
We focus on the two-soliton bound state, and
show that 
it
self-assembles from a variety of initial conditions launched into the SHE: soliton turbulence (as envisaged by Ref.~\citep{zakharov1988soliton}), a single
quasi-coherent solitonic structure, and
two solitons launched on colliding trajectories. This gives us good grounds to believe that 
the statistical attractor that is preferred by the SHE is a multi-soliton bound state, whose two-soliton exemplar we study in detail in this paper.

\subsection{Definition of solitons---the Direct Scattering Transform}
\label{subsec:soli_definition}

At this stage it is necessary to comment on the language used to describe coherent solitary nonlinear wave solutions of PDEs, and in particular solitons. The strictest definition stipulates that solitons are solutions of integrable systems alone. Other definitions categorise solutions of nonintegrable systems that balance nonlinearity against dispersion as solitons~\citep{drazin1989solitonsbook}. For example the SHE has such a solution~\citep{jia2012solitons, horikis2020exact}:
\begin{multline}
\label{eq:SHE_soli}
u(x,t) = \frac{3}{\sqrt{8\beta}} \sech^2 \left[ \frac{1}{\sqrt{4\beta}}(x-s-vt)\right] \\
        \times  \exp\left[iv(x-s)\right] \exp \left[i \left( \frac{1/\beta-v^2}{2} \right) t \right] \exp(i\phi).
\end{multline}
Note that the amplitude, 
and hence the mass (see Eq.~\eqref{eq:waveaction}),
of this solution is set by the nonlocality parameter $\beta$, whereas in the NLSE soliton~\eqref{eq:NLSE_soli} the amplitude is arbitrary\footnote{\label{foot:SHE_soli}
Thus, although sending $\beta\to 0$ in the SHE recovers the NLSE, the solitonic solutions~\eqref{eq:SHE_soli} and~\eqref{eq:NLSE_soli} are  topologically distinct,
in the sense that one cannot continuously transform into the other 
by a continuous change in parameters. It is natural that solutions to the SHE are more restricted than those of the NLSE, due to the introduction of the lengthscale $1/\! \sqrt{\beta}$, which reduces the number of free parameters.}.
As we discuss later, the fixed mass of the SHE's solitonic solution means that it cannot accommodate the mass of an arbitrary initial condition. This implies that the eventual statistical attractor cannot be simply a single structure given by Eq.~\eqref{eq:SHE_soli}.

When they are the only object in the field, 
the solitonic solutions~\eqref{eq:SHE_soli} 
of the SHE
propagate without change of shape, but when they collide they can become distorted, exchange mass, and even merge; we study this in Sec.~\ref{sec:2solicolls}. Some authors refer to solutions such as~\eqref{eq:SHE_soli} loosely as solitons; yet others term these quasi-solitons, see e.g.~\citep{zakharov2004_1DWT}.

In this work, we take the definition of solitons from a pivotal method from integrable systems: the Direct Scattering Transform (DST)~\citep{Zakharov_72, Ablowitz_book, gardner1967method}, a.k.a.\ the nonlinear Fourier transform~\citep{Turitsyn_17, sugavanam_analysis_2019}. This method involves casting a system obeying an integrable equation of motion, in this case the NLSE~\eqref{eq:NLSE}, as an associated linear scattering problem, in which the solution of the equation of motion plays the role of an interaction potential. The solitons present in the system are in one-to-one correspondence with the set of discrete DST eigenvalues obtained from the linear scattering problem. Moreover, the DST eigenvalues are constant in time.
In this sense, one can define a soliton as the physical-space counterpart of a DST eigenvalue. This is especially useful when a field contains many overlapping and interacting solitons, making it hard to associate a soliton with any particular spatiotemporal fluctuation of the field. Such a situation is often referred to as a soliton gas or integrable turbulence~\citep{el2021soliton, suret2024soliton}.

One of the main motivations of this paper is to investigate whether the DST can be used to characterise a system that is nonintegrable, in our case the SHE~\eqref{eq:SHE}, but is nevertheless related to an integrable system. 
We will show that the DST is indeed very useful for the SHE, as it allows one to identify the phenomenology of turbulent processes in the system, and their evolution towards a final statistical attractor.
In keeping with this approach of borrowing the DST from integrable systems, we also borrow the terminology, and speak of solitons as the DST eigenvalues and their physical-space manifestations. We allow ourselves to slip into the looser convention of referring to Eq.~\eqref{eq:SHE_soli} as the SHE soliton. 
Otherwise, we will use phrases such as ``coherent structures'' or ``solitonic waves'' to describe nonlinear waves that are spatiotemporally coherent, but whose profile changes as they move through the system. 
In Sec.~\ref{sec:2solicolls} we describe numerical experiments in which we collide two SHE solitons together. The remnants of these collisions are often two coherent structures that are perturbed versions of the input solitons. For convenience, we will continue to refer to these as solitons, until such time as they merge and form the dominant coherent structure that is the end-state of 
practically
all initial conditions we study in this paper: the two-soliton bound state.

Finally, we will follow the convention of Ref.~\citep{zakharov1988soliton} and continue to refer to chaotic states of a nonintegrable system where there are many strongly nonlinear coherent structures interacting as soliton turbulence. As we will see, such states are indeed characterised by many DST eigenvalues with significant imaginary parts---solitons as we have chosen to define them via the DST.

\section{Numerical methods}
\label{sec:methods}

\subsection{Direct numerical simulation}
\label{subsec:DNS}

In our numerical experiments we solve the SHE, focusing mainly on the case with $\beta=10^{-2}$. We find that this strikes a balance that deviates enough from integrability to access the regime of interest without entirely breaking the correspondence with the NLSE. 
In~\ref{app:1NLS_soli_betascan} we give some details of the dynamics with even smaller $\beta$, for which the system is closer to integrability.

We solve the SHE in a periodic box of length $L=2\pi$, which gives a wavenumber resolution of $\Delta k = 2\pi/L =1$, using a spatial Fourier pseudo-spectral method consisting of $N_x=2048$ Fourier modes~\citep{orszag1969numerical, gottlieb1977numericalBOOK}.
We apply full dealiasing using the $3/2$-rule. To ensure conservation of momentum, we apply dealiasing every time we multiply two fields together in physical space~\citep{krstulovic2011dispersive} (i.e.\ dealiasing happens twice per timestep as the nonlinearity in Eqs.~\eqref{eq:SHE} is cubic). 
Time integration is implemented using the fourth-order exponential time-differencing Runge-Kutta method~\citep{cox2002exponential}, with a timestep $\Delta t$ chosen small enough that the group velocity $v_g=\partial \omega/\partial k$ of the fastest mode is sufficiently resolved, i.e. $\Delta t < \Delta x/\max_k(v_g) = 2L / N_x^2$. 
In all simulations, we closely monitor the waveaction spectrum $n_k \coloneqq |\hat{u}_k|^2 + |\hat{u}_{-k}|^2$, where 
$\hat{u}_k(t) = (1/L)\int_0^L u(x,t)\exp(-ikx)\, dx$, and we see no indication of finite-size effects, in the form of spectral bottleneck, appearing at high $k$. Consequently, we do not add any artificial dissipation to the SHE in the form of hyperviscosity.

To check the convergence of our numerical scheme, we monitor the conservation of the two dynamical invariants of the SHE.
The first invariant is the total waveaction
\begin{equation}\label{eq:waveaction}
N = \int |u|^2 \ dx,
\end{equation}
a.k.a.\ intensity, or 
hereafter the
``mass''.
The second invariant is the
Hamiltonian $H$, 
\begin{equation}\label{eq:hamiltonian}
H = H_2+H_4 =
    \frac{1}{2} \int  \left|\frac{\partial u}{\partial x}\right|^2 \! dx 
    \, - \,
    \frac{1}{2} \int  \left[ \left(1- \beta\frac{\partial^2}{\partial x^2}\right)^{-1/2} \!|u|^2 \right]^2 \!dx,
\end{equation}
where the quadratic energy
$H_2$ is associated with dispersive linear waves, and the quartic energy $H_4$, often called the interaction energy, is associated with strongly nonlinear structures.

We find throughout our simulations that the dynamical invariants $N$ and $H$ are conserved to within $10^{-5}\%$ and $5\times 10^{-5}\%$ respectively.

\subsection{Diagnostics}
\label{subsec:diagnostics}

To help characterise the evolution of the SHE, we utilise two key diagnostics, namely (i) the DST and (ii) the $(k,\omega)$ spectrum, which enable us to extract information about solitonic structures. 
Additionally, we make direct observations of the field, either representing the entire spacetime evolution of $|u(x,t)|$ as a colour plot, or plotting snapshots at fixed times.

\begin{figure}[t]
    \centering
    \includegraphics[width = \linewidth]{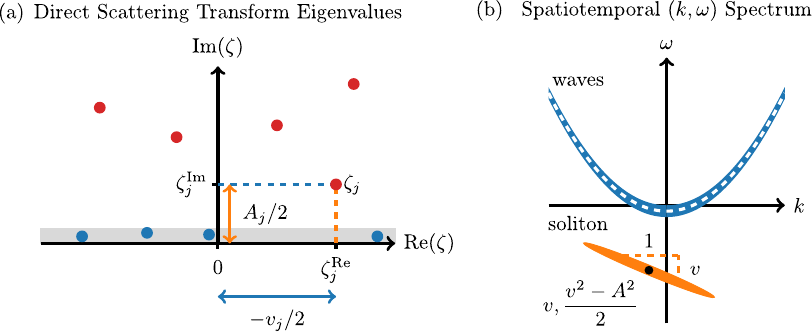}
    \caption{Schematic illustration of the DST and $(k,\omega)$ diagnostics, as applied to an NLSE system. (a) DST output with eigenvalues plotted in the upper half complex plane $\mathbb{C}^+$. Each eigenvalue (red dots) $\zeta_j$ represents a single soliton with an amplitude $A$ equal to $2{\rm Im}(\zeta_j)$ and velocity $v$ equal to $-2{\rm Re}(\zeta_j)$. The grey region illustrates our threshold area denoting spurious eigenvalues produced by the numerical Fourier collocation algorithm. (b) $(k,\omega)$ spectrum, $|\hat{u}(k,\omega)|^2$, with the complex field's wave component (blue area) located around the wave dispersion relation (dashed white curve). Solitons manifest as linear traces orientated with a slope equal to their velocity and centred around the point $(v, -(A^2-v^2)/2)$.
    }
    \label{fig:DST+wk}
\end{figure}

\subsubsection{Direct Scattering Transform}
\label{subsubsec:DST}

As mentioned in Sec.~\ref{sec:intro}, the DST was developed for the integrable NLSE, and consists of recasting the equation as a linear eigenvalue problem, the Zakharov-Shabat problem~\citep{Zakharov_72, Ablowitz_book}. In this problem, the solution $u(x,t)$ of the NLSE at fixed $t$, appears as a parameter. Interpreting the Zakharov-Shabat system as a scattering problem, $u(x,t)$ plays the role of a scattering potential. The solution of the Zakharov-Shabat problem yields a discrete spectrum of complex eigenvalues $\{\zeta_j\}$. Specifying that the corresponding eigenfunctions behave like plane waves at infinity, and imposing their linear independence, generates associated so-called scattering coefficients, whose ratios define a discrete set of norming constants $\{r(\zeta_j)\}$, and also a reflection coefficient $\rho(\xi)$ defined along the real line $\xi\in\mathbb{R}$~\cite{Boffetta_92,Turitsyn_17}. The index $j=1,\dots,m$ labels each soliton, 
with each pair $(\zeta_j,r(\zeta_j))$ containing all the information about the $j$-th soliton in the field $u(x,t)$.  The reflection coefficient is associated with the non-soliton content of the system, i.e., incoherent, 
dispersive waves.

For integrable systems, the discrete eigenvalues remain constant, both in their number (the number of solitons $m$ is conserved throughout the evolution), and in their values $\zeta_j$ (a condition known as isospectrality). In the NLSE the eigenvalues satisfy $\zeta_j^{\rm Im} \coloneqq \mathrm{Im}(\zeta_j) = A_j/2$ and $\zeta_j^{\rm Re} \coloneqq \mathrm{Re}(\zeta_j) = -v_j/2$, where $A_j, v_j$ are the amplitude and velocity of the $j$-th soliton respectively, see Eq.~\eqref{eq:NLSE_soli}, and we have introduced the notation of the real and imaginary parts as $\zeta_j=\zeta_j^{\rm Re}+i\zeta_j^{\rm Im}$. Isospectrality reflects the fact that soliton collisions in the NLSE are elastic.
The norming constants change in time and encode the positions and phase offsets of the solitons. Likewise, the reflection coefficient changes, echoing the evolution of the wave component.

In this work, we apply the DST to our system of interest, the SHE. Tuning the nonlocality parameter $\beta$ away from zero breaks integrability, destroying the precise (in principle, via the Inverse Scattering Transform, invertible) relationship between the solutions of the equation and the DST data. Nevertheless, the DST is still a valid transformation which we can apply to the numerical solutions of the SHE, and can be used as a diagnostic tool to characterise the solutions. The DST is often referred to as the nonlinear Fourier transform~\cite{Turitsyn_17, sugavanam_analysis_2019}.

In particular, we wish to examine coherent quasi-solitonic waves in the SHE. These are large-scale spatial structures, whose support in Fourier space is largest at low $k$. The Fourier transform of Eq.~\eqref{eq:SHE_helmholtz}, 
\begin{equation}
\label{eq:SHE_helm_fourier}
\hat{V}_k = (1+\beta k^2)^{-1} \Big(\widehat{|u|^2}\Big)_k,    
\end{equation}
shows that the SHE-to-NLSE correspondence is best at low $k$, and so we expect that the DST will still yield useful information about the solitonic content of the system.

As our system is spatially periodic, we employ a Fourier collocation method to calculate the DST eigenvalues~\cite{Yang_book_2010}. This method yields only the discrete spectrum $\{\zeta_j\}$, which in the NLSE encodes the amplitudes and velocities of solitons,
and does not generate information about the norming constants $\{r(\zeta_j)\}$, nor the reflection coefficient $\rho(\xi)$. 
The Fourier collocation method generates a fixed number, $2N_x$, of eigenvalues, where $N_x$ is the number of collocation points used. Consequently, the legitimate discrete spectrum is padded with additional spurious eigenvalues with small imaginary parts, due to a shift of the reflection coefficient into the upper half-plane~\cite{agafontsev_bound-state_2023}. As these additional eigenvalues do not correspond to physical solitons, we define a threshold $\zeta^{\rm Im}_{\rm th}$ for the imaginary part of $\zeta_j$.
Eigenvalues with $\zeta^{\rm Im}_j > \zeta^{\rm Im}_{\rm th}$ correspond to realisable solitons, with characteristic widths $\sim \! 1/\zeta^{\rm Im}_{j}$ significantly smaller than the width of the periodic box $L$. 
Conversely, eigenvalues with $\zeta^{\rm Im}_j \ll \zeta^{\rm Im}_{\rm th}$ must be interpreted with caution, as the corresponding solitons are not physically realisable within the box.
Subsequently, we set this threshold such that an NLSE soliton~\eqref{eq:NLSE_soli} whose eigenvalue has $\zeta^{\rm Im}=\zeta^{\rm Im}_{\rm th}$, has a full-width half-maximum equal to $L/4$.
We order the eigenvalue indices $j=1,\ldots,2N_x$ by the size of their imaginary parts, i.e.\ $\zeta^{\rm Im}_1 \geq \zeta^{\rm Im}_2 \geq \ldots$; as discussed above, this corresponds to ordering the solitons by their amplitudes.
  
A schematic representation of the DST output on a 5-soliton system can be seen in Figure~\ref{fig:DST+wk}(a). Here, eigenvalues are represented as red dots, with a position in the upper half complex plane that determines their velocity and amplitude. Spurious eigenvalues are shown as those inside the threshold region depicted in grey (see also \ref{app:DSTnoisefloor}).

\subsubsection{$(k,\omega)$ spectrum}
\label{subsubsec:k-om}

The spatiotemporal, or $(k,\omega)$, spectrum is obtained by taking the double Fourier transform of the field $u(x,t)$ with respect to both $x$ and $t$, to produce 
\begin{equation*}
\hat{u}(k,\omega) = (1/LT)\int_{t-T/2}^{t+T/2} \int_0^L \! u(x,t')e^{(i\omega t'-ikx)} \, dx \, dt'.     
\end{equation*}
(The Fourier transform in time is taken over a time window $T$ long enough to resolve the smallest frequencies of interest.
Data is acquired at a fast enough rate that any temporal artifacts, such as Gibbs ringing due to non-periodicity, manifest at frequencies higher than those of the physical features that we wish to resolve, described below.)
Plotting $|\hat{ u}(k,\omega)|^2$ as a function of $k$ and $\omega$ shows the power in the Fourier coefficient of the spatiotemporal basis function $\exp(ikx-i\omega t)$~\cite{nazarenko2006wave, diLeoni2015spatio}. 

One can then identify features corresponding to the different dynamical entities, see Figure~\ref{fig:DST+wk}(b)
for a schematic representation. Indeed, the $(k,\omega)$ spectrum allows for a full decomposition of the field into wave and solitonic components.
Weak waves are characterised by $(k,\omega)$ distributions concentrated close to the linear wave frequency dispersion relation $\omega(k) = k^2/2$. Nonlinear effects lead to a broadening of the dispersion relation (indicated by the width of the blue parabola), and a vertical shift due to self-interaction (white dashed line).

An NLSE soliton appears in the $(k,\omega)$ spectrum as a linear trace centred at position $(k,\omega)=\left(v , -(A^2-v^2)/2\right)$. This can be seen directly from the exponential factors in Eq.~\eqref{eq:NLSE_soli} where the factors involving $v$ originate from Galilean invariance, and the term involving $A^2/2$ is the effective chemical potential of the soliton solution. The exponential involving $x$ describes the profile of $u(x,t)\in\mathbb{C}$ twisting around the real axis with wavenumber equal to $v$, and the $t$ exponential describes the soliton solution rotating globally in the complex plane.
Carrying out the spatiotemporal Fourier transform of~\eqref{eq:NLSE_soli} shows that the soliton trace in the $(k,\omega)$ spectrum is a straight line with gradient $v$ with a horizontal profile $\sim \! \sech^2[\pi(k-v)/2A]$. 
Such a trace is represented by the orange structure in Fig.~\ref{fig:DST+wk}(b).
The above is also true for the SHE soliton Eq.~\eqref{eq:SHE_soli}, with the amplitude $A$ fixed by $\beta$, and the horizontal profile $\sim\beta^2 k^2 {\rm csch}^2[\pi\sqrt{\beta}\,(k-v)]$.

\section{Soliton turbulence and the emergence of a bound state}
\label{sec:soliturb}

\begin{figure*}[h!t]
    \centering
    \includegraphics[width=0.99\linewidth]{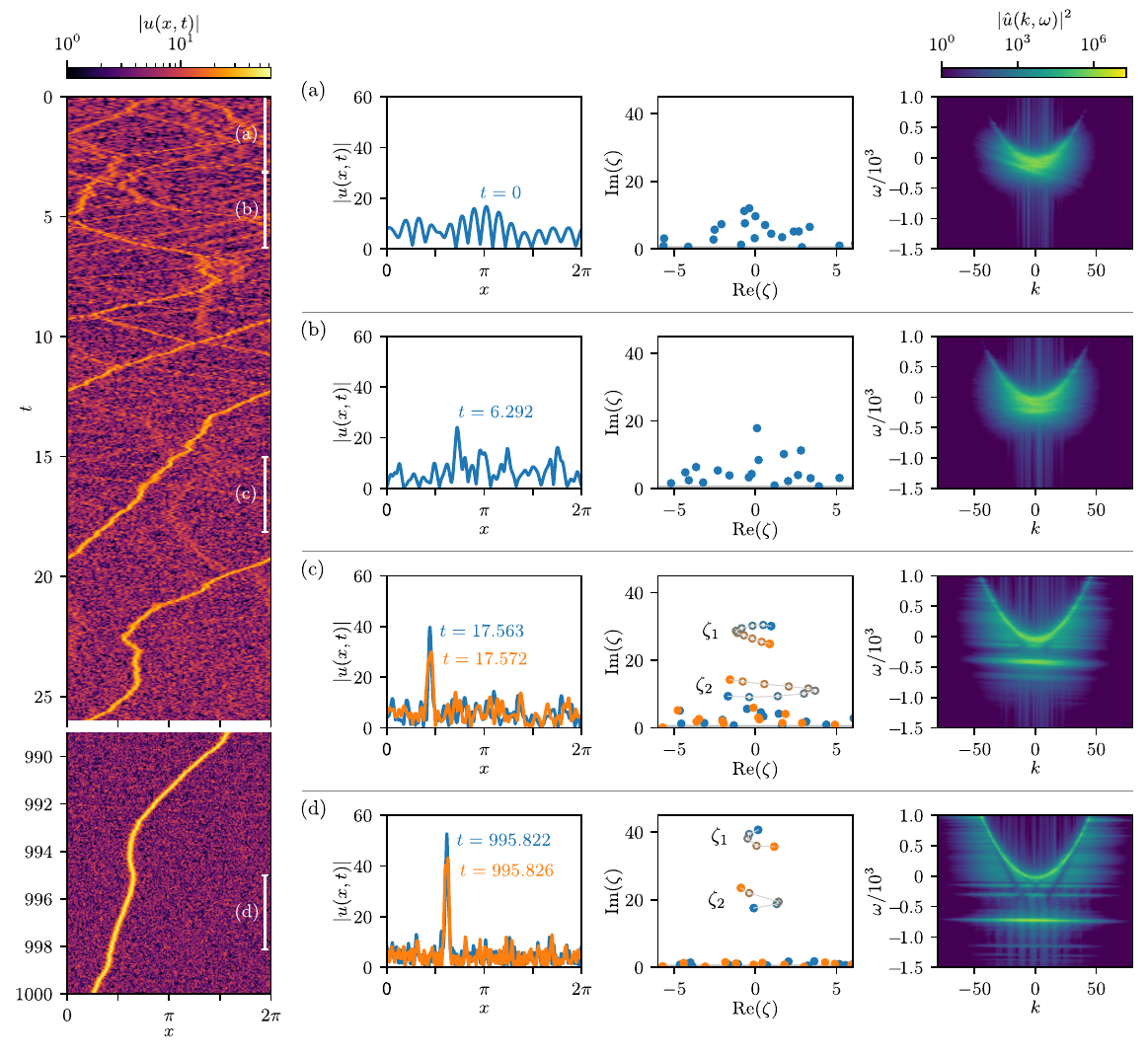}
    \caption{Soliton turbulence leading to a single dominant bound state. Left panel: spacetime evolution of $|u(x,t)|$. White bars (a)-(d) mark representative phases of the evolution detailed in the right panels: (a) random waves developing into (b) soliton turbulence, (c) single dominant solitary bound state embedded in a weakly nonlinear wave field, (d) strengthened bound state and suppressed weak waves. Panels (a)-(d) display: snapshots of $|u(x,t)|$ at the displayed times (first column), DST at the same times (second column), $(k,\omega)$ spectra calculated over a time periods marked by white bars in the left panel (third column).
    In (c) and (d) we also show the primary and secondary DST eigenvalues at intermediate times (unfilled circles).
    \label{fig:soli_turb}}
\end{figure*}

We begin our numerical experiments with a similar initial condition to that studied in~\citep{Laurie_12}, namely a flat-top spectrum of random waves. 
Specifically, we define our initial condition in Fourier space such that $\hat{u}_k(t\!=\!0)=A_0e^{i \theta_k}$ if $k_l\leq |k|\leq k_u$ and zero otherwise, 
where $\theta_k$ is an independent random phase uniformly distributed on $[0,2\pi)$ for each mode $k$. We set the spectral amplitude $A_0$ by specifying the total mass $N$ of the system, using $N= 2 L (k_u-k_l+1)  A_0^2$ (via Eq.~\eqref{eq:waveaction} and using Parseval's identity).

\subsection{Illustrative example of soliton turbulence}
\label{subsec:main_sim_ST}

To illustrate the main finding of this paper, we generate a random state of initial waves in a narrow-band spectrum of modes at large scales by choosing $k_l=6$ and $k_u=9$, and choose a relatively high mass $N=400$. The evolution of this system is shown in Fig.~\ref{fig:soli_turb}, where in the left panel we show the spacetime diagram of $|u(x,t)|^2$ and distinguish four time windows marked by white bars (a)-(d), representing four stages of the dynamics. For each stage (a)-(d), the columns of the right panels show, from left to right: snapshots of $|u(x,t)|$ at specified times, the DST spectra at those times, and the $(k,\omega)$ spectra taken over the whole time window. We choose the length of each representative time window (a)-(d) to be $T=\pi$ time units, in order to resolve one full oscillation of a wave at $k=\Delta k =1$.

\subsubsection{(a), (b) Random waves developing into soliton turbulence}
\label{subsubsec:boundstate_initial_dynamics}
In Fig.~\ref{fig:soli_turb}(a) we show the initial condition: a linear superposition of random waves. As the initial mass is relatively large, the corresponding DST spectrum consists of several eigenvalues that lie significantly above the threshold $\zeta^{\rm Im}_{\rm th}$, i.e.\ the initial condition already contains solitons (in the nonintegrable sense, defined by the existence of these eigenvalues). 

When launched from this initial condition, the spatial fluctuations in the $|u(x,t)|$ field evolve. Soon, solitonic structures emerge in the field and start to stay spatiotemporally coherent, changing in amplitude as they interact, and overlapping significantly in their tails. We term this phase soliton turbulence, in keeping with Ref.~\citep{zakharov1988soliton}.

Meanwhile, the DST eigenvalues ``swarm'' in the complex plane, undergoing  excursions in their real and imaginary parts. This is in complete contrast to the behaviour of DST eigenvalues in the NLSE, which remain constant in time due to integrability of the system: nonintegrability of the SHE breaks isospectrality. The movement of the DST eigenvalues is orchestrated with the dynamics of the field structures. Close examination of this movement strongly suggests that eigenvalues with imaginary parts well above $\zeta^{\rm Im}_{\rm th}$ are associated with individual solitonic structures in the field. 
Recalling that in the NLSE the imaginary part of a DST eigenvalue is proportional to the amplitude of the associated soliton, we see that the link between $\zeta^{\rm Im}$ and amplitude is qualitatively preserved in the solitonic structures observed in the SHE (although the constant of proportionality is no longer 2).

Video animations of the field,  the DST, $(k,\omega)$, and $n_k$ spectra can be found in the Supplementary Material. Video 1 shows the evolution of these diagnostics over time period $t=0$-$30$, which encompasses windows (a)-(c), and Video 2 shows  $t=970$-$1000$ which encompasses window (d).

The details of the dynamics are complex and intricate, but the overall tendency is for the 
high-amplitude coherent solitonic structures to grow at the expense of the lower-amplitude ones
as they collide inelastically.
Occasionally, the colliding solitons merge. As a result,
a small and decreasing number of stronger solitons emerge out of the turbulence and compete for dominance. 
Concomitantly, these interactions manifest as wide excursions of the eigenvalues, and often the identity of the eigenvalue with the largest imaginary part, $\zeta_1$, switches as a result of the interactions. The general tendency is for a few eigenvalues gradually move upwards in the complex plane while the others migrate towards the real axis. In  Fig.~\ref{fig:soli_turb}(b) we show a snapshot of the field and DST spectrum at the end of time window (b).

The emergence of coherent solitonic structures is also evident in the $(k,\omega)$ spectra, as blurry, rectilinear features below the parabolic dispersion relation.The blurriness is due to the structures changing direction during the time window over which the $(k,\omega)$ spectrum is taken.
Between Fig.~\ref{fig:soli_turb}(a) and (b) the dominant coherent structure grows in amplitude and changes velocity, and so its respective trace in the $(k,\omega)$ spectrum moves downwards and changes its slope.

\subsubsection{(c), (d) Emergence and consolidation of a dominant bound state}
\label{subsubsec:boundstate_emergence_colsolidation}

At around $t\simeq 11$, one coherent structure emerges from the turbulent field, standing significantly above the other structures, and making quasiperiodic oscillations in amplitude. 
We choose time window (c) to be representative of the system soon after the dominant structure establishes itself, and window (d) to represent to the system after a long period of evolution, when the structure is well consolidated. 
In the first panel of Fig.~\ref{fig:soli_turb}(c) and (d), we superimpose snapshots of the $|u(x,t)|$ field at respective maxima (blue) and minima (orange) of one oscillation.
In the second panel we show the DST spectra at the times of the chosen maxima and minima (filled blue and orange circles, respectively). We can distinguish two eigenvalues, $\zeta_1$ and $\zeta_2$ by the size of their imaginary parts; they stand well above the others in the complex plane.

At this point, we recall that in Zakharov and Shabat's seminal paper~\citep{Zakharov_72}, they presented a solution of the NLSE consisting of two DST eigenvalues with the same real part, with the two solitons spatiotemporally coincident. Each soliton is trapped in the potential created by the other, and hence this solution is known as a bound state. In physical space, the bound state solution oscillates in amplitude as it propagates, although the DST eigenvalues remain constant due to isospectrality. 

In Secs.~\ref{subsubsec:boundstate_oscillations} and~\ref{subsubsec:boundstate_reconstructions} we present evidence that the dominant coherent structure, which emerges spontaneously out of the soliton turbulence of the SHE, is represented by the pair of distinguished eigenvalues $\zeta_1$ and $\zeta_2$. 
We therefore propose that this coherent structure is a nonintegrable version of a Zakharov-Shabat bound state. The key difference is that, isospectrality being broken, the eigenvalues of the SHE bound state are observed to oscillate. We show this oscillation in Fig.~\ref{fig:soli_turb}(c) and (d), where the positions of $\zeta_1$ and $\zeta_2$ are shown at times in between the maxima and minima (unfilled circles). We discuss these oscillations in detail in Sec.~\ref{subsubsec:boundstate_oscillations}

From $t\simeq 11$ to $t\simeq 40$, the bound state grows overall in amplitude, while the other field fluctuations are suppressed (this is also visible in the colours of the spacetime diagram). The period of oscillations also decreases. Likewise, in the DST spectrum both $\zeta^{\rm Im}_1$ and $\zeta^{\rm Im}_2$ grow, while the subdominant eigenvalues move ever closer to the real axis. 
These observations demonstrate that the bound state strengthens and consolidates by absorbing the other coherent waves through a gradual sequence of inelastic collisions; a ``winner-takes-all'' process where the dominant structure clears out the field around it. The consolidation of the bound state is also evident in the $(k,\omega)$ spectra, where it is visible as a strong, rectilinear trace below the dispersion relation, reminiscent of the $(k,\omega)$ trace of an NLSE soliton, see Fig.~\ref{fig:DST+wk}. 
While the bound state consolidates, this trace moves in the negative $\omega$ direction. This agrees qualitatively with understanding gleaned from the NLSE soliton~\eqref{eq:NLSE_soli}: that the (negative) frequency of a solitonic structure grows as its amplitude increases (however, the bound state arising from soliton turbulence cannot be interpreted as an NLSE soliton, see Sec.~\ref{subsubsec:boundstate_oscillations}).

Another feature that is evident in the $(k, \omega)$ spectrum is a number of fainter rectilinear traces: a pair of secondary sidebands that flank the primary trace, a series of subdominant sidebands, and decorations of the dispersion relation at each integer value of $k$.
In all simulations we have analysed, as the bound state changes velocity, all these rectilinear features change in slope and remain parallel, indicating that they are associated with the dominant bound state.

Regarding the decorations of the dispersion relation, we recall that our wavenumber resolution is $\Delta k=1$, so the discrete scarring is likely a finite-size effect. We conjecture that in a physical system, the result of this decoration at every continuous wavenumber would be a general broadening of the dispersion relation. We interpret this broadening as an interaction of the bound state with 
weakly nonlinear
waves at every frequency.

Additionally, we note existence of a primary and secondary set of furrows in the $(k,\omega)$ spectrum that cut the rectilinear traces lying above the primary trace. We currently lack any explanation of these gaps in excitation.

We have observed that around $t\simeq 40$, the growth of the bound state saturates, after which the diagnostics are qualitatively similar to what is shown in Figs.~\ref{fig:soli_turb}(d) and~$\ref{fig:soli_turb_oscillations}$. During this time the bound state undergoes periods of acceleration, deceleration, and changes of direction, presumably by exchanging momentum with the residual background waves~\citep{zakharov1988soliton}, akin to the Brownian motion of a particle suspended in a fluid.

\subsubsection{Examination of the bound state---oscillations of eigenvalues}
\label{subsubsec:boundstate_oscillations}

\begin{figure}[t!]
    \centering
    \includegraphics[width=0.99\linewidth]{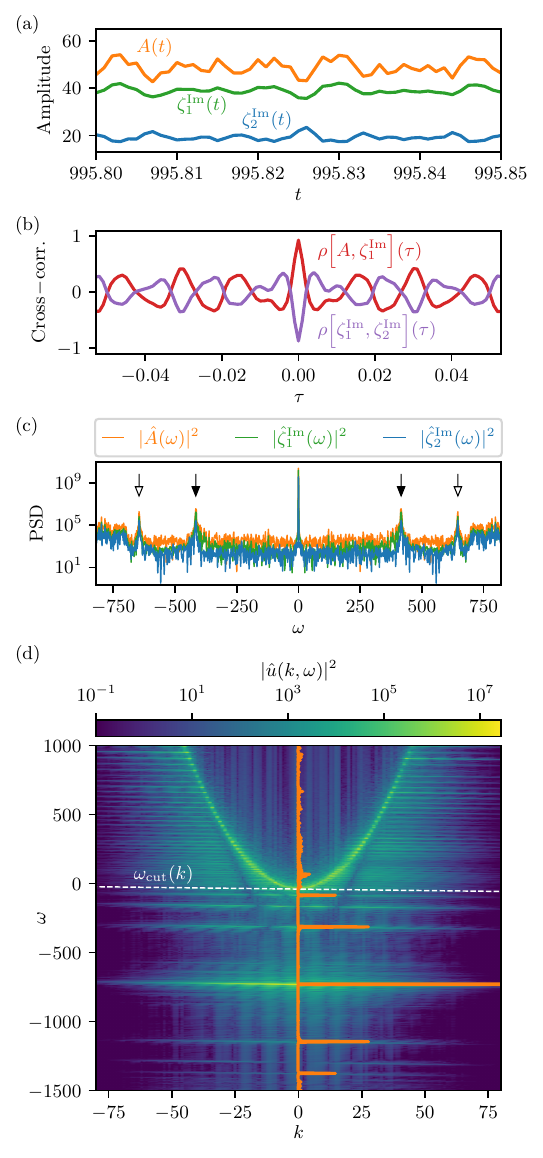}
    \caption{
    Examination of the consolidated bound state shown in Fig.~\ref{fig:soli_turb}(d). 
    (a) Comparison of the timeseries $A(t)$, $\zeta^{\rm Im}_1(t)$, and $\zeta^{\rm Im}_2(t)$.
    (b) Normalised 
    cross-correlations
    $\rho[A, \zeta^{\rm Im}_1](\tau)$ and $\rho[\zeta^{\rm Im}_1, \zeta^{\rm Im}_2](\tau)$. 
    (c) Temporal PSDs $|\hat{A}(\omega)|^2$, $|\hat{\zeta}^{\rm Im}_1(\omega)|^2$, and $|\hat{\zeta}^{\rm Im}_2(\omega)|^2$. All three PSDs have a primary peak at $\omega = 0$, secondary peaks at $\omega \simeq \pm 415.8$ (filled arrows), and tertiary peaks at $\omega \simeq \pm  645.7$
    (unfilled arrows).
    (d) $(k,\omega)$ spectrum. A primary, a pair of secondary, and a sequence of subdominant solitonic traces are observed below the parabolic dispersion curve of weak waves. The PSD $|\hat{A}(\omega)|^2$ from (b) is superimposed in orange, centred on the primary soliton trace at $(k,\omega)\simeq(-0.212,-730)$ (see main text). The white dashed line $\omega_{\rm cut}(k)$ divides the wave component above from the solitonic component below, see Sec.~\ref{subsubsec:boundstate_reconstructions} and Fig.~\ref{fig:ST_chop_uhat_DST}.
\label{fig:soli_turb_oscillations}}
\end{figure}

Returning to the oscillations of the consolidated bound state, and the DST eigenvalues $\zeta_1$ and $\zeta_2$ that comprise it, in Fig.~\ref{fig:soli_turb_oscillations}(a) we plot timeseries of the peak amplitude $A(t) \coloneqq \max_x(|u(x,t)|)$ (orange), $\zeta^{\rm Im}_1$ (green), and $\zeta^{\rm Im}_2$ (blue), for a representative time interval within time window (d). All three quantities appear to oscillate with the same fundamental frequency. The fluctuations between $A(t)$ and $\zeta^{\rm Im}_1$ seem strongly positively correlated, whereas the fluctuations of $\zeta^{\rm Im}_1$ appear negatively correlated with those of $\zeta^{\rm Im}_2$. 

To confirm this, in Fig.~\ref{fig:soli_turb_oscillations}(b) we show the normalised 
cross-correlations
$\rho[A, \zeta^{\rm Im}_1]$ (red) and $\rho[\zeta^{\rm Im}_1, \zeta^{\rm Im}_2]$ (purple), calculated over the whole time window (d), 
where the normalised cross-correlation between 
timeseries $f(t)$ and $g(t)$, assumed statistically stationary, is
\begin{equation*}
    \rho[f,g](\tau) = \frac{\int_{-\infty}^{\infty} \left(f(t+\tau)-\mu_f\right) \left( g^*(t)- \mu_g^*\right) dt}{\sigma_f \sigma_g},
\end{equation*}
and $\mu_f$, $\sigma_f$ respectively denote the empirical mean and standard deviation of $f(t)$, etc. There is a positive peak of $\rho[A, \zeta^{\rm Im}_1](\tau)$ and a negative peak of $\rho[\zeta^{\rm Im}_1, \zeta^{\rm Im}_2](\tau)$ at lag $\tau=0$. This demonstrates conclusively that the bound state amplitude $A$ and $\zeta^{\rm Im}_1$ oscillate in synchrony, whereas $\zeta^{\rm Im}_1$ and $\zeta^{\rm Im}_2$ oscillate in exact phase opposition. We have also calculated $\rho[\zeta^{\rm Re}_1, \zeta^{\rm Re}_2]$ and find a strong negative peak at $\tau=0$, demonstrating that the real parts of these eigenvalues likewise oscillate 
in phase opposition.
The 
phase-opposed  
oscillation of $\zeta_1$ and $\zeta_2$ is clearly visible in their trajectories, shown in Fig.~\ref{fig:soli_turb}(c) and (d).

In Fig.~\ref{fig:soli_turb_oscillations}(c) we plot the temporal power spectral density (PSD) of the timeseries that are partially shown in (a). Namely, we plot $|\hat{A}(\omega)|^2$ (orange),  $|\hat{\zeta}^{\rm Im}_1(\omega)|^2$ (green), and $|\hat{\zeta}^{\rm Im}_2(\omega)|^2$ (blue). In each of these PSDs, we observe a primary peak at $\omega = 0$, representing the mean value of the signal, 
and a pair of secondary peaks (indicated by filled arrows) at $\omega\simeq\pm 415.8$, the fundamental oscillation frequency of the eigenvalues. Tertiary peaks (unfilled arrows)  are also seen at $\omega\simeq\pm 645.7$.
The symmetry around $\omega = 0$ is due to the time series being real-valued. 

To interpret the peaks of the PSDs, in Fig.~\ref{fig:soli_turb_oscillations}(d) we again plot the $(k,\omega)$ spectrum, and superimpose the PSD $|\hat{A}(\omega)|^2$, aligning its primary peak on $(k,\omega)=(-0.213, -730)$, the approximate centre of the primary solitonic trace (see below). 
We do this to take into account the fact that $A(t)$ is the evolution of the bound state's peak amplitude, which is real-valued. Its PSD gives frequency information about the fluctuations in the height of the peak. However, the whole profile is rotating in the complex plane. This rotation frequency is 
detected in
the $(k,\omega)$ spectrum as 
it
is obtained by Fourier transforming the complex field $u$ directly.
Thus, aligning the $\omega=0$ peak of the PSD with the $\omega$ position of the primary trace amounts to transforming into the frame corotating with the primary soliton in the complex plane, and studying frequencies relative to the rotation frequency. We see that the secondary peaks of the PSD align perfectly with the secondary solitonic traces in the $(k,\omega)$ spectrum, showing that both the secondary PSD peaks and the secondary $(k,\omega)$ traces contain information about fluctuations of the amplitude of the bound state. These fluctuations are mirrored, with the appropriate phase shifts, in 
$\zeta^{\rm Im}_1$ and $\zeta^{\rm Im}_2$.

Regarding the $(k,\omega)$ placement of the PSD, we choose the horizontal position to be the measured average velocity of the bound state $\langle v \rangle \simeq -0.213$ during this time window,
where angle brackets denote time averaging.
We expect the relationship $k=\langle v \rangle$ to be preserved as this comes directly from the Galilean invariance of the SHE, see Sec.~\ref{subsubsec:k-om}. This position agrees with the centre of the soliton trace, with the caveat that the bound state has a velocity below the $k$ resolution of our system.
The vertical placement of the PSD is chosen empirically as the maximum of the primary solitonic trace on the vertical line $k=\langle v \rangle$, i.e.\  $\Omega = \max_\omega(|\hat{u}(k,\omega)|^2)_{k=\langle v \rangle} \simeq-730$. This frequency agrees with neither soliton equation~\eqref{eq:NLSE_soli} nor~\eqref{eq:SHE_soli}: averaging the bound state amplitude throughout time period (d) gives $\langle A \rangle
\simeq 48.9$. If the primary linear trace corresponded to an NLSE soliton, it would have frequency $\omega=
-(\langle A \rangle^2-\langle v \rangle^2)/2 \simeq - 1194$, which is far from the measured frequency $\Omega\simeq-730$.  Neither does the bound state correspond to a SHE soliton~\eqref{eq:SHE_soli}, which has a fixed amplitude $3/\!\sqrt{8\beta}=10.6$ in our case of $\beta=10^{-2}$, corresponding to frequency $-(1/\beta-\langle v\rangle^2)/2 \simeq -100$.
Nor does the SHE bound state frequency correspond to the frequency of a two-soliton bound state of the NLSE~\cite{Zakharov_72}, which for our normalisation of~\eqref{eq:NLSE} is $|\omega^{\rm NLSE}_{\rm bs}| = 2|(\zeta^{\rm Im}_1)^2-(\zeta^{\rm Im}_2)^2|$. Using the time-averaged eigenvalues of our SHE bound state, this would correspond to $|\omega^{\rm NLSE}_{\rm bs}|\simeq 2305$, which compares very poorly to the measured frequency $|\Omega| \simeq 730$.

Thus, we conclude that the bound state arising from SHE soliton turbulence can be modelled neither as an NLSE soliton, nor a SHE soliton,
nor an NLSE bound state.

\begin{figure}[t]
    \centering
    \includegraphics[width=0.99\linewidth]{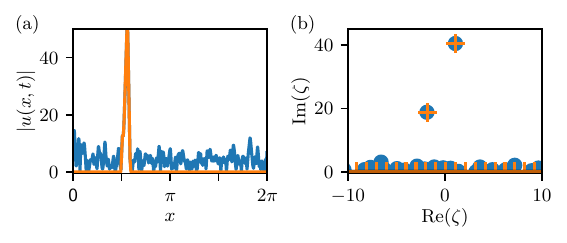}
    \caption{(a) full $|u(x,t)|$ field (blue) and spatially filtered bound state $|u_{\rm bs}(x,t)|$ (orange) at time $t \simeq 996.5713$. (b) DST of the full field (blue points) and filtered component (orange plusses).
    \label{fig:ST_chop_absu_DST}}
\end{figure}

\subsubsection{Examination of the bound state---reconstruction of system components}
\label{subsubsec:boundstate_reconstructions}

To confirm the identity of the dominant coherent wave as a bound state, we filter out the surrounding wave field spatially by setting $u(x,t)=0$ outside the coherent structure, from the first local minima either side of $\max_x(|u|)$ that 
satisfy $|u(x,t)| < \langle |u(x,t)|\rangle_x$ 
(with $\langle \ldots \rangle_x$ denoting spatial averaging), 
giving the filtered bound state $u_{\rm bs}(x,t)$. Its absolute value is shown in orange in Fig.~\ref{fig:ST_chop_absu_DST}(a), together with that of the original field, $|u(x,t)|$, in blue. In (b) we show the DST spectrum calculated from the full (blue points) and filtered (orange plusses) field. We see perfect coincidence of $\zeta_1$ and $\zeta_2$, while all subdominant eigenvalues of the filtered field are practically zero. This definitively identifies the dominant coherent wave arising out of soliton turbulence as a bound state comprising of $\zeta_1$ and $\zeta_2$. 
Filtering the bound state from the waves in this way allows us to calculate the proportion of the total mass that the bound state accumulates: $\int_0^L \! |u_{\rm bs}|^2 dx/\int_0^L \! |u|^2 dx \simeq 0.618$.

\begin{figure}[t]
    \centering
    \includegraphics[width=0.99\linewidth]{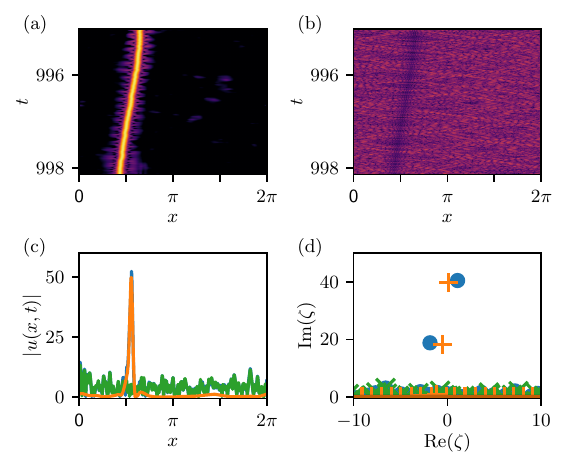}
    \caption{
    $|u(x,t)|$ field and DST, reconstructed from filtered components of $\hat{u}(k,\omega)$.
    (a) Spacetime diagram of the bound state, reconstructed by band-passing the soliton traces in $\omega \leq \omega_{\rm cut}(k)$ (see Fig.~\ref{fig:soli_turb_oscillations}(d)).
    (b) Spacetime diagram of the background waves, reconstructed from the dispersion relation in $\omega > \omega_{\rm cut}(k)$.
    (c) $|u(x,t)|$ snapshots taken from the reconstructed spacetime diagrams at $t \simeq 996.5713$. Orange: snapshot of the bound state taken from (a), green: wave component taken from (b), blue (mostly obscured): original full field.
    (d) DST spectra of the snapshots shown in (c), respectively shown in orange plusses, green crosses, and blue points.
    \label{fig:ST_chop_uhat_DST}}
\end{figure}

Further confirmation comes from band-pass filtering the doubly Fourier transformed field $\hat{u}(k,\omega)$ to select either the weakly nonlinear wave
component, 
or the solitonic components. Examining the $(k,\omega)$ spectrum Fig.~\ref{fig:soli_turb_oscillations}(d), the 
waves
can be separated 
from the solitons
by the line $\omega_{\rm cut}(k) = \langle v \rangle k - 40$, shown by the white dashed line (the downshift of the dispersion relation by $40$ is due to the nonlinear correction to the linear wave frequency, which is towards negative $\omega$ since the SHE resembles the focusing NLSE at low $k$).
The wave component consists of the parabolic dispersion relation lying in $\omega > \omega_{\rm cut}(k)$. Filtering out the waves and taking the double inverse Fourier transform of the $\omega \leq \omega_{\rm cut}(k)$ component of the $\hat{u}(k,\omega)$ field yields the spacetime diagram in Fig.~\ref{fig:ST_chop_uhat_DST}(a). The trajectory of the bound state is recovered exactly, with almost no waves in the field.
Conversely, band-passing the Fourier field in $\omega > \omega_{\rm cut}(k)$ and inverting recovers Fig.~\ref{fig:ST_chop_uhat_DST}(b): the wave component, with the field suppressed at the bound state trajectory. This shows that the $(k,\omega)$ spectrum is a discriminating tool to separate the wave from the solitonic components of a system. 

Taking snapshots of the field at the temporal mid-point $t \simeq 996.5713$ of the spacetime plots Fig.~\ref{fig:ST_chop_uhat_DST}(a) and (b), we recover the isolated bound state (orange) and wave (green) components of the $|u(x,t)|$ field shown in (c). The original snapshot at this time is shown in the background in blue, which is almost completely obscured by the snapshots of the two reconstructed field components. Figure~\ref{fig:ST_chop_uhat_DST}(d) shows the DST spectra of the original snapshot (blue points), reconstructed bound state component (orange plusses), and reconstructed wave component (green crosses). The eigenvalues $\zeta_1$ and $\zeta_2$ are reasonably well recovered, with clear separation from the eigenvalues representing the wave component, near the real axis. This evidence cements the mutual link between the bound state, its representation below the dispersion relation in the $(k,\omega)$ spectrum, and its DST spectrum which consists of the two eigenvalues $\zeta_1$ and $\zeta_2$.

\subsection{Phase diagram of the bound state attractor}
\label{subsubsec:phasediag}

The results of Sections~\ref{subsubsec:boundstate_initial_dynamics} and~\ref{subsubsec:boundstate_emergence_colsolidation}
are typical of many simulations we have run starting from a spectrum of large-scale random waves. 
The self-assembly of a bound state from random turbulence is a robust phenomenon that occurs generally, provided the initial spectrum of such waves is in some sense large enough. However, for low-amplitude initial conditions, no long-lived coherent structure emerges.
To characterise this more fully, we set up initial flat-top spectra of random waves at three different placements of $k_l$ and $k_u$, keeping the spectral width the same. Specifically, we set up initial spectra supported on $|k|\in[2,5], [6,9],$ and $[10,13]$. We define $k_0 = (k_l+k_u)/2$, a characteristic wavenumber for each initial spectrum. For these three placements of the initial condition, we vary the total mass $N$. We launch these initial spectra into the SHE and evolve the system until a stable bound state emerges in the system, or up to a long computational time of $t=12000$.  
(To assist
with
the speed and memory requirements
of this parameter scan, we drop the spatial resolution to $N_x=1024$. Energy and mass are still well conserved, and the $n_k$ spectrum shows no sign of a spectral bottleneck.)

\begin{figure}[t]
    \centering
    \includegraphics[width=0.99\linewidth]{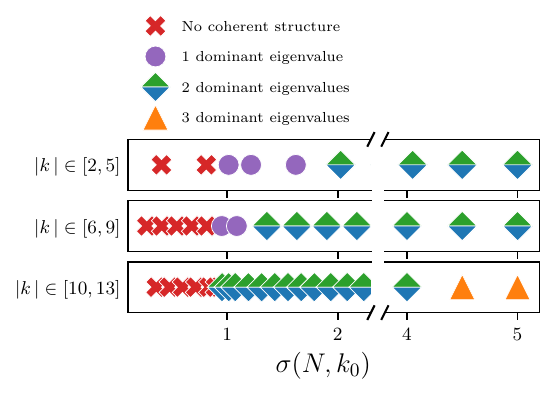}
    \caption{Qualitative phase diagram of the final states reached from an initial flat-top random wave spectrum, supported on the given wavenumber intervals. The control parameter $\sigma(N,k_0)$, given by Eq.~\eqref{eq:sigma_control_param}, is a proxy for $|H_4/H_2|$ of the initial condition.
\label{fig:Soli_turb_phase_diag}}
\end{figure}

To determine a control parameter from characteristics of the initial spectrum that predicts whether or not a bound state will appear, we note once more that solitonic structures balance linear dispersion with 
self-interaction.
The initial condition should therefore be nonlinear enough to allow a solitonic structure to form, i.e.\ we should expect these structures when 
$|H_4/H_2| \gtrsim 1$
in the initial condition. 
Noting Eq.~\eqref{eq:hamiltonian} for the expressions of these energy contributions, we form the non-dimensional quantity 
\begin{equation}
\label{eq:sigma_control_param}
\sigma(N,k_0) = \frac{N}{Lk_0(1+\beta k_0^2)}.
\end{equation}
This quantity is a proxy for $|H_4/H_2|$ that contains the characteristic mass and wavenumber of the initial spectrum, but which does not depend on the particular realisation of the initial condition.

Figure~\ref{fig:Soli_turb_phase_diag} summarises the results of these simulations, in the form of a qualitative phase diagram, where we note 
four
kinds of emergent states. 
Red crosses indicate that no long-lived coherent structures emerge from the turbulence. Here, the waveaction spectrum evolves from the initial condition towards a broadband spectrum of random waves, and if any coherent structures appear during the evolution they persist only transiently. 
Purple
circles indicate a low-amplitude oscillating coherent structure where $\zeta^{\rm Im}_1$ is at least $1.5\zeta^{\rm Im}_2$ on average, but $\zeta_2$ does not rise significantly above the other eigenvalues (i.e.\ different eigenvalues in turn rise above the others to transiently assume the role of $\zeta_2$ before dropping towards the real axis again). We conjecture that this structure is also in fact a bound state, with an amplitude too low for $\zeta_2$ to emerge clearly.
Green-blue diamonds denote the formation of a larger-amplitude bound state, with both $\zeta_1$ and $\zeta_2$ standing well above all other eigenvalues in the complex plane. The simulation reported in Sec.~\ref{subsec:main_sim_ST}
is of this type. 
Orange triangles denote a final state with three dominant eigenvalues with a clear ordering $\zeta^{\rm Im}_1 > \zeta^{\rm Im}_2 > \zeta^{\rm Im}_3 > \zeta^{\rm Im}_j$ for subdominant eigenvalues $j\neq 1,2,3$. Again, the coherent structure presents as a single oscillating solitary wave in physical space: the system organises into a three-soliton bound state.

Figure~\ref{fig:Soli_turb_phase_diag} demonstrates that $\sigma(N,k_0)$ is indeed 
a good control parameter---it predicts the formation of a bound state for $\sigma \gtrsim 1$, and no coherent structure for $\sigma \lesssim 1$, for each of the three placements of the initial spectrum.

Around the transition at $\sigma\approx 1$, we observe slowing down of the dynamics, in that for a given placement of the initial spectrum, the bound state takes progressively longer to form as $\sigma\to 1^+$. In several realisations close to but above the transition, we also observe transient fragility of the bound state: a coherent structure initially forms with a certain amplitude, then it weakens, before consolidating again and saturating at a higher amplitude.
It is of course possible that these will collapse into random waves eventually, but we do not observe any further weakening for as long as we continue the simulations.

Likewise, for $\sigma\to 1^-$, we sometimes see a coherent structure emerging transiently only to disperse again, associated with one eigenvalue rising above the others in the DST spectrum, before eventually collapsing back into the ``swarm'' of eigenvalues near the real axis.

For values  of $\sigma$ significantly below the transition, long simulations show no sign of the random waves strengthening into coherent structures. It may be the case that the formation of the bound state is delayed beyond our simulation time. However, since the control parameter $\sigma$ is physically motivated by energy considerations, we do not expect coherent structures ever to form for initial conditions below the threshold.

For simulations around $\sigma \approx 1$, the observation of coherent structures transiently emerging then dispersing indicates a competition between the formation of solitonic structures ($H_4$ dominating), and their disruption by energetic linear waves ($H_2$ dominating).
This lends a heuristic interpretation to the ``squeezing out'' of the low-amplitude bound state with only one dominant soliton (purple circles in Fig.~\ref{fig:Soli_turb_phase_diag}) from parameter space when the initial condition is concentrated at high characteristic wavenumber $k_0$. For a given mass $N$ of an initial condition, increasing $k_0$ increases the energy $H_2\propto \sum_k k^2 |\hat{u}_k|^2$ associated with linear waves, and so any incipient solitonic structures are more strongly disrupted by the energetic waves. When $\sigma \approx 1$, the energy balance allows the bound state to remain stable against perturbation. However since for large $k_0$ we have $\sigma \sim N/k_0^3$, this requires a much larger $N$ than for small $k_0$, and so when the bound state is robust enough to persist, it has a large enough amplitude for the second eigenvalue to rise above the ``swarm'' of subdominant eigenvalues.

The orange triangles in the bottom-right corner of Fig.~\ref{fig:Soli_turb_phase_diag} indicate a second transition from a two-soliton to a three-soliton bound state. Increasing $\sigma$ towards this portion of the phase diagram from below leads to progressively larger $\zeta^{\rm Im}_1$ and $\zeta^{\rm Im}_2$, until the third dominant eigenvalue $\zeta_3$ rises above the ``swarm'' of subdominant eigenvalues, oscillating with the same frequency as $\zeta_1$ and in phase opposition. 
We note that systems in this portion of the phase diagram are initialised with high $N$. Evidently, the amount of mass that each soliton comprising the bound state has an upper limit, leading to a third soliton growing from the ``swarm'' to absorb the mass of the initial condition at some critical value.

This clearly raises the possibility that the two-soliton bound state is the low-amplitude member of a family of bound states with an arbitrary number of dominant solitons. We focus on the two-soliton structure for the rest of this paper, and leave a full characterisation of the full family of multi-soliton bound states and their transitions to future work.

The classification criteria of the final states reported in Fig.~\ref{fig:Soli_turb_phase_diag},
and the boundaries of the various phases,
is admittedly somewhat subjective. We present it as a first attempt to categorise the final states that appear 
out
of soliton turbulence in the SHE, and leave it to future work
to detail a more principled and quantitative classification scheme,
in particular to determine the parameters that control the transition to a bound state with three or more dominant solitons.

\subsection{Summary}
\label{subsec:boundstate_summary}
To summarise the findings above, we have observed the spontaneous self-assembly of a single coherent, dominant, solitonic wave, emerging out of soliton turbulence in the SHE. 
In the system of Sec.~\ref{subsec:main_sim_ST} that we have studied closely, this
coherent structure is a bound state comprised of two solitons, defined and detected by the Zakharov-Shabat DST eigenvalue spectrum. As the bound state propagates, its amplitude oscillates periodically, as do its constituent solitons. The oscillations of the solitons' DST eigenvalues are 
in phase opposition in both
their real and imaginary parts, corresponding to oppositional fluctuations in their amplitudes and velocities. It is natural to interpret these anti-correlated fluctuations in velocity as both solitons in the potential well created by the other, analogous to a binary star system orbiting a common barycentre. Likewise, we interpret the 
phase-opposed
oscillations in amplitude as the solitons exchanging mass back and forth as they propagate.

The findings of Sec.~\ref{subsubsec:phasediag} indicate that the appearance of the bound state is a robust phenomenon that occurs in the SHE as long as the system contains enough interaction energy to form coherent structures.
It appears that as the mass of the initial condition increases, the system self-organises into a bound state comprising of progressively more solitons to accommodate the majority of this mass.
We therefore propose that the pulsating
multi-soliton bound state is a statistical attractor of the SHE. 

The initial condition we have considered, a flat-top spectrum of random waves, is sufficiently general that we expect the emergence of a bound state from any other class of initial condition. Put otherwise, the self-organisation of random waves into a bound state represents ``order emerging from chaos''. More coherent initial conditions would also readily evolve towards the universal attractor: ``order emerging from order''. The rest of this paper describes results we have obtained in this direction.

\section{Single NLSE soliton propagation in the SHE}
\label{sec:1soli}

\begin{figure}[t]
    \centering
    \includegraphics[width = \linewidth]{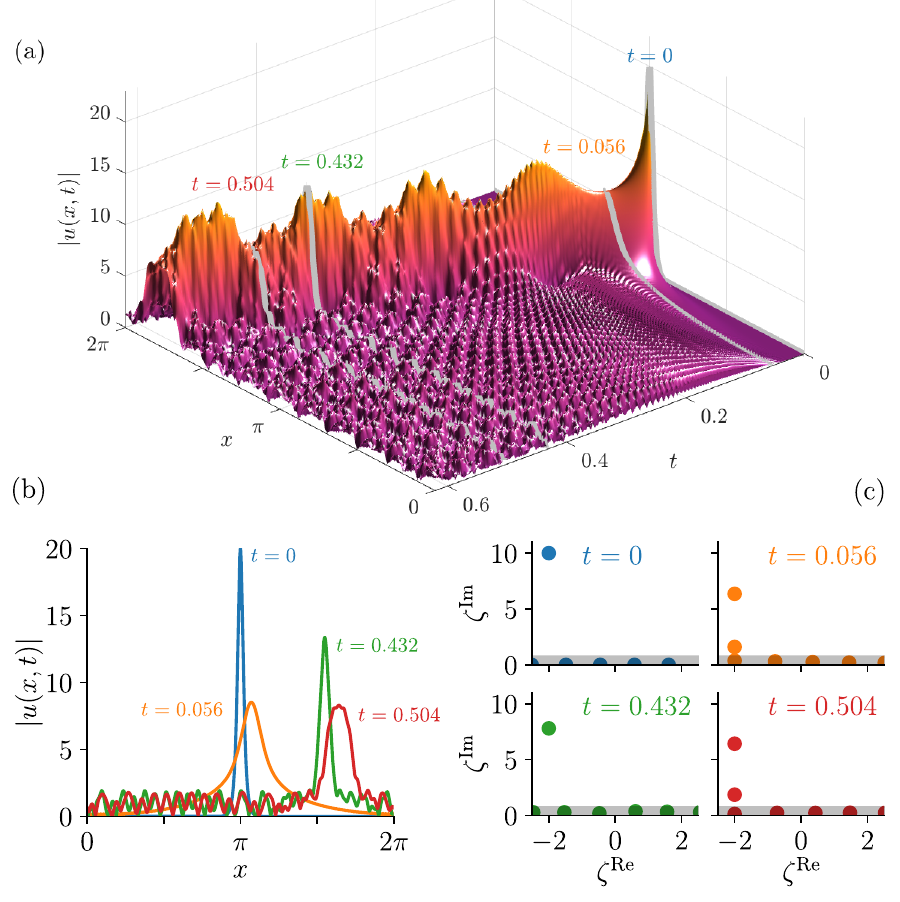}
    \caption{(a) Spacetime evolution of $|u(x,t)|$ for an NLSE soliton launched into the SHE. Snapshots of the field at the displayed times are plotted in (b) (marked in (a) by grey lines). (c) DST spectra at the corresponding times. \label{fig:1soli_xt}}
\end{figure}

In Sec.~\ref{sec:soliturb} we carried out a thorough examination of the final bound state created by a turbulent process, involving the interaction of many solitons that are initially present in the system.
We now demonstrate that a similar bound state can arise directly when we launch a nonlinear wave that is different to the exact soliton naturally supported by the SHE. 

We launch a single NLSE soliton~\eqref{eq:NLSE_soli} with parameters $(A,v,s,\phi)=(20,4,\pi,0)$, into the system. Since $\beta=10^{-2}\neq0$, we do not expect this initial condition to propagate in the SHE without a change of profile. Indeed, in Fig.~\ref{fig:1soli_xt} we see that the coherent structure immediately emits waves into the domain while the profile relaxes: the peak of the structure falls while its width broadens. This initial relaxation is followed by a rebound, and thereafter the amplitude and width of the coherent structure oscillate periodically. These oscillations become noisy once the radiated waves travel across the periodic domain and re-encounter the structure.
This is seen in Fig.~\ref{fig:1soli_xt}(a), where we show the spatiotemporal evolution of the $|u(x,t)|$ field, with snapshots at times $t=0$ (initial condition), $t=0.056$ (minimum of the first relaxation), $t=0.432$ (maximum of a subsequent noisy oscillation), and $t=0.504$ (ensuing minimum) shown in (b). The  DST spectra at the corresponding times are shown in (c). Once again, we see that the DST eigenvalue $\zeta_1$ oscillates in phase with the peak of the coherent structure.


After the initial condition has relaxed, and as the residual coherent structure starts to rebound and oscillate, the eigenvalue $\zeta_2$ grows from the real axis and oscillates above and below the threshold $\zeta^{\rm Im}_{\rm th}$, in phase opposition with $\zeta^{\rm Im}_1$. Both eigenvalues have the same constant real part 
$\zeta^{\rm Re}_1=\zeta^{\rm Re}_2\simeq -2.00$,
indicating that the two solitons corresponding to these eigenvalues travel at the same speed, not fluctuating like in the case of Sec.~\ref{sec:soliturb}. 
The velocity of the coherent structure remains identical to the initial soliton velocity, measured to be $v \simeq 4.00$ over any time interval during the simulation, so we drop the angle brackets. The relation between $v$ and $\zeta^{\rm Re}_{1,2}$ is in perfect agreement with the Zakharov-Shabat theory for NLSE solitons.

\begin{figure}[t]
    \centering
    \includegraphics[width=0.99\linewidth]{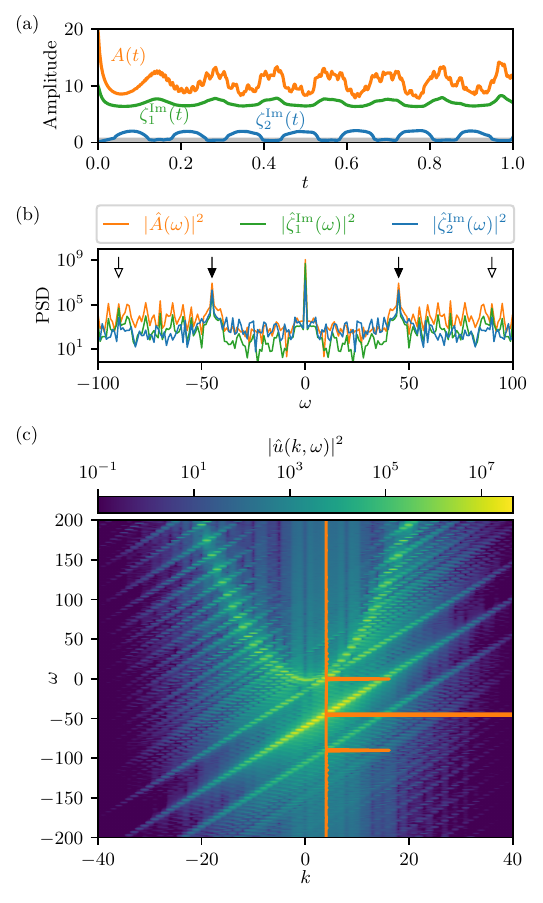}
    \caption{
    As per Fig.~\ref{fig:soli_turb_oscillations}(a), (c) and (d), for a single NLSE soliton \eqref{eq:NLSE_soli} with $A=20$ and $v=4$, launched into the SHE. In (b), the temporal PSDs have secondary peaks at $\omega \simeq 45.0$; tertiary peaks at $\omega \simeq 90.0$ are evident in the PSDs  $|\hat{\zeta}_1^{\rm Im}(\omega)|^2$ and $|\hat{\zeta}_2^{\rm Im}(\omega)|^2$, and not prominent in $|\hat{A}(\omega)|^2$. In (c) we overlay the PSD of $A(t)$ onto the $(k,\omega)$ spectrum at $(4,-45)$.
    \label{fig:1soli_oscillations}}
\end{figure}

We repeat the analysis of Sec.~\ref{subsubsec:boundstate_oscillations} and display the results in Fig.~\ref{fig:1soli_oscillations}, showing: (a) timeseries of the peak amplitude $A(t) \coloneqq \max_x(|u(x,t)|)$, and the DST eigenvalues' imaginary parts $\zeta^{\rm Im}_1(t)$, and $\zeta^{\rm Im}_2(t)$, (b) the temporal PSDs of $A(t)$ and $\zeta^{\rm Im}_1$(t), and (c) the $(k,\omega)$ spectrum, overlaid with the PSD of $A(t)$. Carrying out a cross-correlation study as before, we once again find that the peak amplitude  and $\zeta^{\rm Im}_1$ oscillate in phase, and that $\zeta^{\rm Im}_1$ and $\zeta^{\rm Im}_2$ oscillate 
in exact phase opposition. 
For brevity we omit displaying this study.

Just like in the case of Sec.~\ref{sec:soliturb}, where a bound state emerged from a period of soliton turbulence, Figs.~\ref{fig:1soli_xt} and~\ref{fig:1soli_oscillations} demonstrate that when the NLSE soliton is launched into the SHE, the system again self-organises into a bound state comprising of two oscillating eigenvalues. In this case the bound state has a lower amplitude, such that the secondary soliton $\zeta_2$ fluctuates above and below the threshold $\zeta^{\rm Im}_{\rm th}$ set by the domain size; physically it dips in and out of existence.

Looking in detail at the $(k,\omega)$ spectrum, in Fig.~\ref{fig:1soli_oscillations}(c) we have centred the PSD of $A(t)$ horizontally at $k=v=4$, which according to Eqs.~\eqref{eq:NLSE_soli} and~\eqref{eq:SHE_soli} should be the centre of the $(k,\omega)$ primary soliton trace. Inspection of the figure shows this to be the case. 
As for the vertical positioning, we centre the PSD on $\Omega = \max_\omega (|\hat{u}(k,\omega)|)_{k=v} = -45.0$. 
This compares very favourably to the frequency $\omega = -47.1$ of an NLSE soliton with amplitude $\langle A \rangle
=10.5$ and velocity $v=4$. 
(We note that a two-soliton bound state of the NLSE whose eigenvalues had imaginary parts $\langle \zeta^{\rm Im}_1\rangle$ and $\langle \zeta^{\rm Im}_2\rangle$ would have frequency $|\omega^{\rm NLSE}_{\rm bs}|\simeq 93.7$, which is very far from the measured $|\Omega|$.)

Placing the PSD at $(k,\omega) \simeq (4, -45.0)$, we find that the secondary peaks of the PSD align perfectly with the secondary solitonic $(k,\omega)$ traces, as was the case in Sec.~\ref{sec:soliturb}. Furthermore, we find that the centre of the upper secondary trace is at $\omega \simeq 0$ to within numerical resolution, i.e.\ the oscillations in the amplitude and eigenvalues have the same frequency as the rotation of the whole profile in the complex plane, $\Omega$. We also observe that the secondary trace here is tangent to the dispersion relation of linear waves. 
To model this, let us 
first
ignore the spatial profile, and transform into the comoving frame with velocity $v$. The leading temporal behaviour is the rotation of a complex amplitude $A$, i.e.\ $u\sim A e^{-i\Omega t}$. Next, we include sinusoidal oscillations of the amplitude about its average 
$\langle A \rangle
$, i.e.\ $A\to \langle A \rangle
+ 2\Delta A \cos(\Omega t)$. Expressing the cosine as a sum of complex exponentials, we immediately see that the temporal variation of such an amplitude-modulated solitary wave is $u \sim 
\langle A \rangle
e^{-i\Omega t} + \Delta A + \Delta Ae^{-2i\Omega t}$, a signal that rotates in the complex plane with frequency $-\Omega$, and that has weaker sidebands at frequencies $0$ and $-2\Omega$. Dressing these signals with a solitonic profile that is linear in the $(k,\omega)$ plot reproduces exactly what we observe in Fig.~\ref{fig:1soli_oscillations}(c).

\subsection{Discussion---mechanism for the secondary soliton's creation}
\label{subsec:1soli_discussion}

To conclude this Section, we propose the following heuristic explanation of our observations of one initial NLSE soliton launched into the SHE. Since the profile~\eqref{eq:NLSE_soli} no longer balances dispersion with self-focusing in the SHE, the initial condition releases waves and relaxes. As the deviation from integrability is in some sense small (however see~\ref{app:1NLS_soli_betascan}), the nonlinear wave remains mostly coherent, and its initial velocity is unperturbed by this collapse. The remaining coherent nonlinear wave has a soliton component as it is detectable in the DST spectrum as the eigenvalue $\zeta_1$.

Since we do not damp the waves, they recirculate in the system, with each wave packet travelling at its own group velocity. 
As the coherent structure passes through the wave field, the two interact nonlinearly. We speculate that this wave-structure interaction amounts to a forcing of the waves by the coherent structure at every $k$, leading to the broadening of the dispersion relation, observed as the decorations of the parabola in Fig.~\ref{fig:1soli_oscillations}(c) (and indeed in Fig.~\ref{fig:soli_turb_oscillations}(d) in Sec.~\ref{sec:soliturb}). The bright spot where the secondary soliton trace meets the dispersion relation tangentially suggests that this forcing is most efficient where the coherent structure is resonant with the waves, namely where the primary soliton's velocity and the wave group velocity are equal.
(We note that the resonant excitation of waves by solitons was reported in a nonintegrable Korteweg-De Vries equation~\citep{osborne1998soliton}.)
We further speculate that this efficient forcing of the resonant waves causes them to grow preferentially. Eventually they become nonlinear enough to undergo a modulational instability, which creates the secondary soliton, represented in the DST by $\zeta_2$. The primary and secondary solitons are spatially coincident, the resulting structure being the bound state. 
As the system evolves and the bound state propagates, the two solitons periodically exchange mass, leading to phase-opposed oscillations in the imaginary parts $\zeta^{\rm Im}_1$ and $\zeta^{\rm Im}_2$.
We note that while this scenario is physically plausible, it remains somewhat speculative. Further work is necessary in order to put it on a more sound mathematical footing.

In addition to the results presented here, we have also launched simulations initialised with a single NLSE soliton with a variety of amplitudes and velocities. Furthermore, we have made runs starting from a SHE soliton, and from an NLSE or SHE soliton amplified vertically by a factor $\gamma \in [0.70, 2.0]$, 
in the manner of
Ref.~\citep{kuznetsov_nonlinear_1995}. In each case we see qualitatively the same dynamics, further reinforcing the idea that the generation of the second soliton via the wave-primary soliton interaction is generic, and that the resulting two-soliton bound state is a universal solution favoured by this nonintegrable system. 

It is natural to assume that this mechanism of resonant interaction with waves is active in the initial transient phase of soliton turbulence examined in Sec.~\ref{sec:soliturb}. We conjecture that during the initial phase as coherent structures are assembling, each such structure accumulates waves into its own secondary soliton, so that the coherent structures that finally merge into the final dominant bound state are each themselves bound states. We turn to the question of collisions and  mergers of coherent structures next.

\section{Collisions of SHE solitons}
\label{sec:2solicolls}

\begin{figure*}[t]
    \centering
    \includegraphics[width=0.99\linewidth]{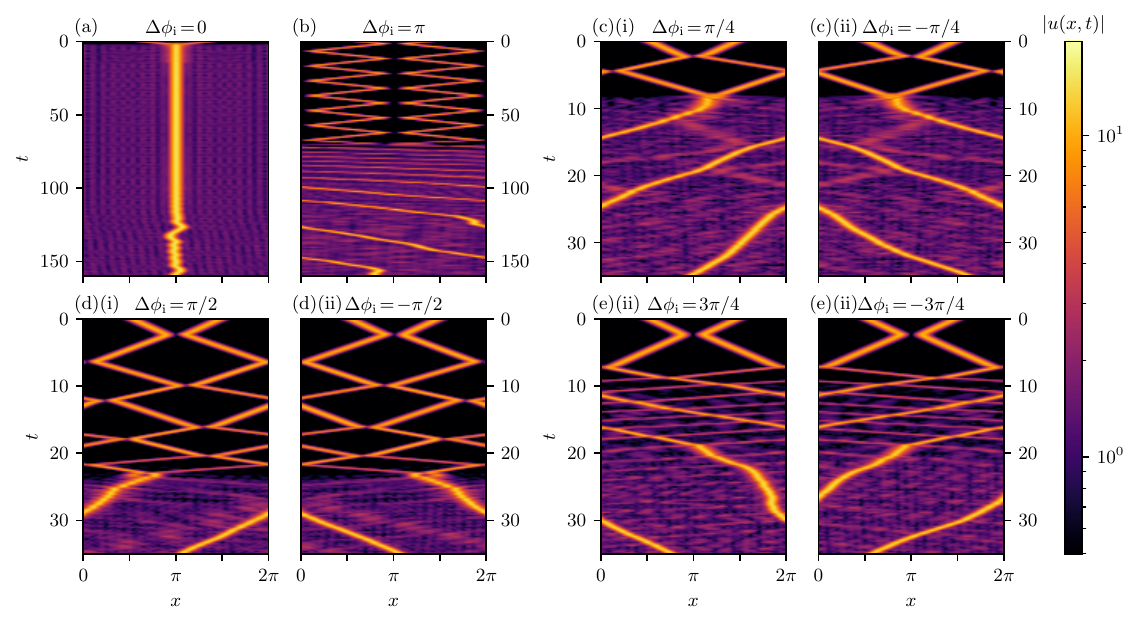}
    \caption{Spacetime evolution of two SHE solitons colliding in our periodic computational domain. The solitons are initialised with the displayed initial phase differences $\Delta\phi_{\rm i}=\phi_2-\phi_1$. In each case the two initial solitons merge into a single bound state.    
    In (c)-(e) the respective subplots (i) and (ii) retain the reflection--phase inversion symmetry of the initial condition, until the symmetry is spontaneously broken by the soliton merger event. Thereafter the phase-space trajectories of (i) and (ii) diverge. Note the difference in timescale between (a), (b) and (c)-(e).}
    \label{fig:2SHE_solis_dPhi_scan}
\end{figure*}

In this Section, we examine a key feature that separates the dynamics of  coherent solitonic waves
in nonintegrable systems to those of solitons in integrable systems: the ability of coherent waves to undergo inelastic collisions and mergers. We observe such events happening frequently during the initial phase of soliton turbulence (see Fig.~\ref{fig:soli_turb}(a) and (b) and Video 1), en route to forming the dominant bound state of the system.

Here we study the collision and merger processes in a cleaner environment, allowing us to determine some necessary conditions for two solitonic waves merge into a single bound state. 
We initialise the system with the linear sum of two SHE solitons~\eqref{eq:SHE_soli} with positions $s_1 \!=\! L/4$, $s_2 \!=\! 3L/4$, and velocities $v_1 \!=\! 0.5$, $v_2 \!=\! -0.5$, and study the effect of varying the initial the phase difference $\Delta \phi_{\rm i} = (\phi_2-\phi_1)|_{t=0}$. 

The periodic boundary conditions mean that the two solitons that we launch into the system will cycle through the domain and collide with each other many times. 
In contrast to integrable dynamics, the solitons perturb each other at every encounter, meaning that they do not retain their profiles after the first collision. 
From this point, in order to not overburden the narrative we will use the term soliton to refer not only the initial profiles that we launch into the system, but also to the perturbed remnants that emerge after each collision. As we will see, this comports with the terminology we have already established, of solitons being objects that are represented by physically-relevant eigenvalues in the DST spectrum, but here we will mainly be concerned with their spatiotemporal manifestations. 

Figure~\ref{fig:2SHE_solis_dPhi_scan} shows the spacetime plots of our simulations, for SHE solitons with initial phase differences of (a) $\Delta\phi_{\rm i} = 0$, (b) $\pi$, (c) $\pm \pi/4$, (d) $\pm \pi/2$, and (e) $\pm 3\pi/4$. We group simulations (c), (d), and (e) this way in order to visualise the initial symmetry under reflection and phase inversion $\{x\to L-x, v\to-v, \Delta\phi(t) \to-\Delta\phi(t) \}$.

In all cases the two solitons eventually merge into a single dominant coherent structure remaining in the system, surrounded by incoherent weak waves which are mainly emitted following the merger event. Using the same methods as presented in Secs.~\ref{sec:soliturb} and~\ref{sec:1soli}, we find once again that the final coherent structure is a two-soliton bound state, with a primary DST eigenvalue that oscillates in phase with the peak amplitude, and 
in phase opposition to
the secondary eigenvalue, and with a $(k,\omega)$ spectrum consisting of a primary and two solitonic traces, which align perfectly with the primary and secondary peaks of $|\hat{A}(\omega)|^2$. This lends further credence to the bound state being the final statistical attracting state of the SHE.

We also note that the time at which the two initial solitons merge to form the final bound state depends on $\Delta\phi_{\rm i}$. In the case of $\Delta\phi_{\rm i}=0$ the bound state forms on the first collision of the input solitons. By contrast, for $\Delta\phi_{\rm i}=\pi$ the single bound state emerges after the solitons have recirculated through the system and collided many times. 
For intermediate values of $\Delta\phi_{\rm i}$, the merger of the two initial solitons occurs after an intermediate number of collisions (note the different timescales shown in Fig.~\ref{fig:2SHE_solis_dPhi_scan}(a) and (b), vs.\ (c)-(e)). 
As the solitons are launched with identical initial amplitudes and velocities, they rotate in the complex plane with the same initial frequency $\Omega=(1/\beta-v^2)/2$. Therefore, they retain their initial phase differences as they approach the first collision.

\begin{figure}[t]
    \centering    \includegraphics[width=0.99\linewidth]{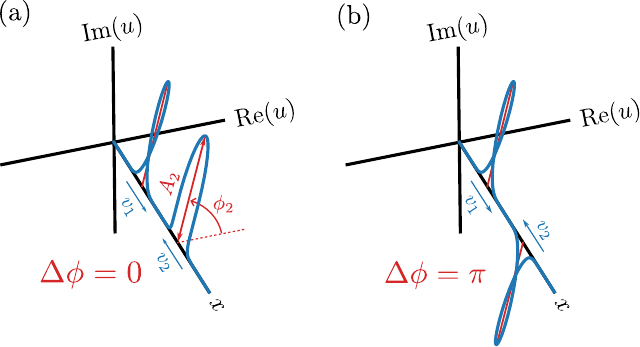}
    \caption{SHE solitons approaching each other before their first collision, with phase differences $\Delta \phi=0$ (a), and $\Delta\phi=\pi$ (b). In (a) we demonstrate schematically how we identify the phase and amplitude of one of the solitons. As time evolves, the solitons approach each other with velocities $v_1$ and $v_2$, and the solitons rotate in the complex plane clockwise, with frequency $\Omega_j \sim (A_j^2-v_j^2)/2$.
    \label{fig:x_uRe_uIm}}
\end{figure}

This dependence on the phase difference can be explained by noting that if two solitons approaching each other are to merge, the merger is a highly nonlinear process. If the solitons approach with phase difference $\Delta\phi = 0$, as shown in Fig.~\ref{fig:x_uRe_uIm}(a), it is natural to assume that their amplitudes will add in the complex plane, leading to a large nonlinearity. This favours the solitons merging into a single coherent structure.
Conversely, if two solitons approach with phase difference $\Delta\phi=\pi$, Fig.~\ref{fig:x_uRe_uIm}(b), their complex amplitudes will tend to cancel out, the nonlinearity will be small, and their merger is inhibited.

\begin{figure}[t]
    \centering    \includegraphics[width=0.99\linewidth]{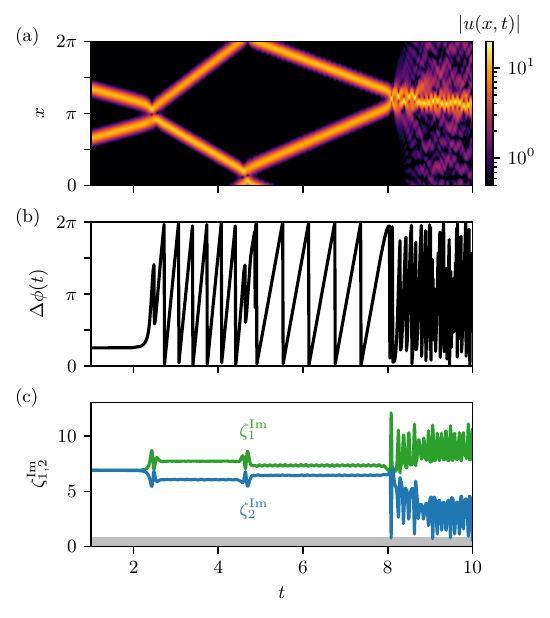}
    \caption{Initial collisions, and merger at $t\simeq8.05$, of two SHE solitons initialised with $\Delta\phi_{\rm i}=\pi/4$. (a) spacetime diagram, (b) evolution of the phase differences $\Delta\phi(t)$ as found via the first and second local maxima of $|u|$ (see text and Fig.~\ref{fig:x_uRe_uIm}), evolution of $\zeta^{\rm Im}_1$ (green) and $\zeta^{\rm Im}_2$ (blue).  
    \label{fig:2SHE_dPhi_study_1pi__4}}
\end{figure}

These considerations suggest that solitons with a general phase difference merge when their phases synchronise. We confirm this in Fig.~\ref{fig:2SHE_dPhi_study_1pi__4}, where we show the first two quasi-elastic collisions of the solitons, followed by the merger at the third collision, for the case of $\Delta\phi_{\rm i}=\pi/4$. 
The spacetime plot in (a) shows the general evolution of $|u(x,t)|$. In (b) we plot the evolution of the phase difference between the two solitons. This is found at each timestep by first finding the $x$ positions of the two largest local maxima of $|u(x,t)|$ to detect the soliton peaks. We then take the phase of the solitons as the arguments of $u(x,t)$ at these positions, as shown in Fig.~\ref{fig:x_uRe_uIm}(a) for $\phi_2$. In (c) we plot the evolution of $\zeta^{\rm Im}_1$ (in green) and $\zeta^{\rm Im}_2$ (in blue). 
We see that $\Delta\phi(t)$ remains $\pi/4$ until the first collision. During this collision, some mass is exchanged from one soliton to the other,
with the change in soliton amplitudes reflected in $\zeta^{\rm Im}_1$ increasing and $\zeta^{\rm Im}_2$ decreasing. The difference in amplitudes means that the rotation frequencies of the solitons $\Omega_j\sim(A_j^2-v_j^2)/2$ are now different, leading to a linear growth of $\Delta\phi$. Another exchange of mass (and consequently the soliton rotation frequencies) happens at the second collision. At the third collision at $t\simeq8.05$, the phase difference approaches $2\pi$, i.e.\ the solitons are nearly synchronised in phase, and the solitons merge into a bound state. (The evolution of $\Delta\phi$ after this point loses its interpretation as the phase difference between the two solitons.) 
Note that before the merger, $\zeta_1$ and $\zeta_2$ are each associated 
with
a different soliton, but after the merger they are the primary and secondary eigenvalues of the bound state.
During the merger event, the larger soliton appears to capture the smaller one. As the solitons merge, the identity of $\zeta_2$ changes several times. Namely, the old $\zeta_2$ drops into the grey area below $\zeta_{\rm th}^{\rm Im}$, with a different eigenvalue arising from the grey area to assume the new identity of $\zeta_2$. 
After a short transient phase, the bound state stabilises and $\zeta_1$ and $\zeta_2$ retain their identity, oscillating in phase opposition as we have observed before.
These dynamics can be seen in Video 3 of the Supplementary Material.

We have repeated the study above for all cases in Fig.~\ref{fig:2SHE_solis_dPhi_scan}. For $\Delta\phi_{\rm i} =\pm \pi/4$ and $\pm\pi/2$ it is clear that the binary soliton mergers occur when the solitons collide with a phase difference close to 
$0 \bmod{2\pi}$.
In the case of $\Delta\phi_{\rm i} =\pm 3\pi/4$ and $\pi$, a large number of mass-exchanging collisions occur before the eventual merger. Either soliton can gain mass at the expense of the other, but the general tendency is for the larger-amplitude soliton to accrete mass from the smaller. This leads to a pre-merger condition where the soliton amplitudes are very different, and hence the phase is evolving very rapidly. The timescales of the merger and phase evolution become comparable, so it is hard to associate the merger with one particular instance that $\Delta\phi \approx 0$. The hypothesis of binary solitons mergers being associated with phase synchronisation is nevertheless consistent with these cases. Additionally, it is natural to assume that it is easier for colliding solitons of very different sizes to merge, because it is hard for a much smaller soliton to escape the potential created by a comparatively large soliton.

We have also repeated the study of binary soliton mergers, launching two identical NLSE solitons into the SHE, and scanning over initial phase differences. Again, every initial condition leads to a single bound state, with mergers of the initial solitons happening quickly for $\Delta\phi_{\rm i}=0$, delayed for $\Delta\phi_{\rm i}=\pi$, and at intermediate times for intermediate phase differences. We find clear merger events when solitons collide with aligned phases, as well as cases where mergers are preceded by a large number of mass-exchanging collisions. For a summary of one such study, see~\ref{app:2NLS_solis_in_SHE}.

Finally, we note that in Fig.~\ref{fig:2SHE_solis_dPhi_scan}(c)-(e) the reflection--phase inversion symmetry is broken in every case after the merger occurs, as evidenced by the loss of bilateral symmetry between the respective subfigures (i) and (ii). 
We attribute this symmetry breaking due to numerical effects and the discretised representation of the SHE being a chaotic dynamical system. At the mergers, which are large-amplitude, high-nonlinearity events, the differences in the phase space trajectories of systems (i) and (ii) are amplified and the trajectories diverge exponentially thereafter.
Convergence studies show that increasing the spatial resolution $N_x$ or decreasing the timestep $dt$ cause the reflection--phase inversion symmetry to be retained for longer after the merger, before the trajectories diverge. Crucially, increasing the spatial or temporal resolution does not influence the time of merger, indicating that this qualitative feature of the dynamics is robust. As for Fig.~\ref{fig:2SHE_solis_dPhi_scan}(a) and (b), the breaking of bilateral symmetry in both of these cases is a numerical artifact that can be delayed by increasing the resolution. We expect that if we were to realise infinite precision, the case with $\Delta\phi_{\rm i}=\pi$ would never result in a merger. We conjecture that this would be the only case in which the binary soliton merger could be avoided up to arbitrary time. Likewise with infinite precision the system initialised with $\Delta\phi_{\rm i}=0$ would remain symmetric for arbitrarily long times.

\section{Other oscillating solitonic waves discussed in previous literature
}

The identification of the statistical attractor of the SHE as a bound state shares some commonalities with oscillating solitonic structures that have been described in the literature.
As we have noted, the bound state that we realise is the nonintegrable version of the Zakharov-Shabat bound state~\citep{Zakharov_72}. 

The oscillating behaviour of solitons has also been observed experimentally in Ref.~\cite{sugavanam_analysis_2019}, where the integrability of 1D NLSE is broken by the existence of losses in optical fibres and by the pumping of lasers. Here too, solitons with oscillating amplitudes are associated with a pair of DST eigenvalues. 

Related observations were recently made in Refs.~\cite{kachulin2021bound, gelash2024bisolitons}, where two-soliton states (termed bi-solitons by those authors) were constructed in nonintegrable models of two-dimensional hydrodynamics, starting from bi-solitonic states of the NLSE. In their narrow-band limit, these hydrodynamic models are nonintegrable extensions of the NLSE. The bi-solitons discussed there qualitatively resemble those of the SHE that we report on here, in terms of their profile, their oscillatory spacetime dynamics, and in~\cite{gelash2024bisolitons}, the phase-opposed oscillations of their DST eigenvalues. 
This is strong evidence that the two-soliton states that we describe are generic attractors in weakly nonintegrable models, for both spatially nonlocal (in our case) and spatially local (in the hydrodynamic case) nonlinearities. However, we note that the DST spectra reported in Ref.~\cite{gelash2024bisolitons} have eigenvalues whose real parts have opposite sign. Interpreting these as correlates of the soliton velocities, it appears that the bi-solitons reported are in fact counter-propagating solitons recirculating in a periodic box, in contrast to the bound states we describe here. Our results also show that coherent multi-solitonic bound states are generic, self-assembled structures that evolve out of chaotic processes (turbulence and collisions) and do not require careful preparation in order to produce them.

Further back in the literature,
Kuznetsov~{\it et al.}~\cite{kuznetsov_nonlinear_1995} studied an amplified soliton (Eq.~\eqref{eq:NLSE_soli} multiplied by an overall factor) in the NLSE. This amplified soliton showed an initial relaxation, followed by periodic oscillations in amplitude that eventually decayed towards a new equilibrium soliton state at a rate $\sim\! t^{-1/2}$. Their decay was associated with the use of absorbing boundary conditions that dissipated the waves radiated from the initial relaxation. 
In our numerical experiments reported in Sec.~\ref{sec:1soli}, the waves emitted by the relaxation of the initial soliton recirculate within the domain, and the oscillations remain stable and persistent. Indeed, the generation of the secondary soliton, via resonance of the primary soliton with packets of weak waves, relies on waves co-existing with the primary soliton. 
As said in Sec.~\ref{sec:1soli}, we leave it to future work to describe this process more fully and mathematically.

Moreover, Agafontsev~{\it et al.}~\cite{agafontsev_bound-state_2023} studied the forced NLSE, in which a period of forcing created a state with many oscillating solitons with zero velocity. They termed the final state a bound-state soliton gas, which they suggested could be a universal asymptotic state of integrable turbulence. The DST of this universal state has many eigenvalues arranged 
densely
in a line along the imaginary axis. By contrast, in the nonintegrable SHE we find that the universal attracting bound state consists of 
oscillating eigenvalues that are well spaced in the complex plane.

Another class of 
coherent, strongly nonlinear
solutions of the NLSE are breathers: spatiotemporally periodic solutions of the NLSE, which have been studied, for example, in Refs.~\citep{agafontsev2024multisoliton, soto-crespo_integrable_2016}. In the taxonomy of these authors, breathers are distinguished from solitons by their asymptotics: the breather solutions 
asymptotically
tend to a nonzero constant 
as 
$|x|\to\infty$. The theoretical DST spectrum of breathers involves a vertical branch cut in the complex plane to create the nonzero background, with isolated eigenvalues placed in different positions to generate different classes of breather~\citep{agafontsev2024multisoliton}. When the DST is taken numerically, the branch cut is reproduced by a line of closely-spaced eigenvalues. The principal difference with our work is that we consider the evolution of solitonic structures on a zero background. Consequently, a vertical branch cut does not feature in our DST spectra.

\section{Conclusion and perspectives}
\label{sec:discussion}

Through numerical experiments in the SHE, we have demonstrated the existence of a 
multi-soliton bound state, surrounded by weakly nonlinear waves. 
This state is reached by the system from a variety of random or coherent initial conditions, demonstrating that it is an attracting end-state of evolution in the SHE. 
Focusing for the most part on the two-soliton bound state, we have thoroughly characterised it by the use of the DST and $(k,\omega)$ diagnostics.

In addition, by launching coherent structures into the SHE, we have identified basic processes involved with the creation of the secondary soliton of the bound state, and of the collision and merger of solitonic structures. Launching an NLSE soliton into the system, we identify the generation of the secondary soliton as a resonance process between the primary soliton and packets of incoherent weak waves travelling with the same phase velocity. Launching two SHE solitons, we find that in order for them to merge, they must be in phase synchrony when they collide. If they collide with detuned phases, they nevertheless exchange mass. On average this mass transfer is from the smaller soliton to the larger.
We conjecture that the resonant generation of secondary solitons, soliton mergers, and collisions with gradual mass exchange, are all active when the system is launched from a multi-soliton state that evolves into soliton turbulence. The overall effect is for a decreasing number of coherent structures to become increasingly large, until the 
emergence of the
final statistical attractor of a single dominant bound state, surrounded by weak waves.

A crucial feature of the final attracting state of the SHE is that the dominant coherent structure appears capable of absorbing the majority of the mass of the initial condition, which can be arbitrary. 
This arbitrariness means the final state
cannot be a single soliton of the SHE, due to the latter having an amplitude fixed by the nonintegrability parameter $\beta$, 
see Eq.~\eqref{eq:SHE_soli}.
Nor can the dominant structure be a single soliton of the NLSE. Although this soliton can be of arbitrary amplitude, it is not a natural soliton of the system and immediately radiates waves and undergoes a partial collapse, as we 
saw
in Sec.~\ref{sec:1soli}. 
The remedy is for the system to assemble itself into a 
multi-soliton system, in which the solitons exchange mass back and forth periodically.
This indicates that a final oscillating 
multi-soliton bound state may be a feature of nonlocal systems in general, as they possess a natural lengthscale which controls the width and amplitude of their solitons.

The work we have presented here opens up lines of enquiry in a few directions. 
Firstly, it remains to be shown whether the
multi-soliton
bound state is in fact {\it the} universal statistical attractor predicted by previous works~\citep{zakharov1988soliton, Jordan2000meanfield, Rumpf2001coherent},
for weakly nonintegrable systems in general. The
observations we report here certainly suggest that it is a promising candidate.

Secondly, we have offered heuristic explanations of many features of the 
two-soliton
bound state, such as the periodic exchange of mass between the two constituent solitons being responsible for their oscillations in amplitude, the resonant interaction between the primary soliton and waves to generate the bound state, and the merger process requiring phase synchrony between the colliding solitons on energetic grounds. All of these require a better mathematical explanation.

Thirdly, we have demonstrated conclusively that 
understanding the dynamics of a nonintegrable system can be aided by use of the DST of a related integrable system.
This suggests a programme of work to examine other near-integrable systems in this vein. 
Further understanding could be gleaned by using established DST perturbation theory techniques in which the parameter encapsulating the weakness of nonintegrability (in our case $\beta k^2$) is used as a small expansion parameter~\cite{kaup1976perturbation, karpman1977perturbation, kivshar1989dynamics}. Such an approach has already been demonstrated in Ref.~\cite{gelash2024bisolitons}, to examine the trajectories of the DST eigenvalues in hydrodynamic equations related to the NLSE.
Such an approach would be useful in characterising the bound state, but might not be amenable to analysing the long-term dynamics of the system from a turbulent initial condition, as in Sec.~\ref{sec:soliturb}, as it self-assembles into the bound state.

Finally, we have demonstrated that the DST yields information that is fully consistent with the $(k,\omega)$ spectrum, a tool that enjoys widespread use in the study of weak wave turbulence. It is also well known that the DST becomes asymptotically equivalent to the linear Fourier transform in the low-amplitude limit~\citep{Ablowitz_book}. We hope that diagnostic techniques of this kind 
can
be adapted to uncover detailed information about the interactions between coherent structures and weakly nonlinear waves, in systems where both components are present. To do so would make great strides towards developing a self-consistent theory of strongly and weakly nonlinear waves in nonintegrable systems---a long-term objective of the wider theory of wave turbulence~\citep{nazarenko2011book}.

\appendix

\section{Single NLSE soliton launched into the SHE - variation with $\beta$}
\label{app:1NLS_soli_betascan}

\begin{figure}[t]
    \centering
    \includegraphics[width=0.9\linewidth]{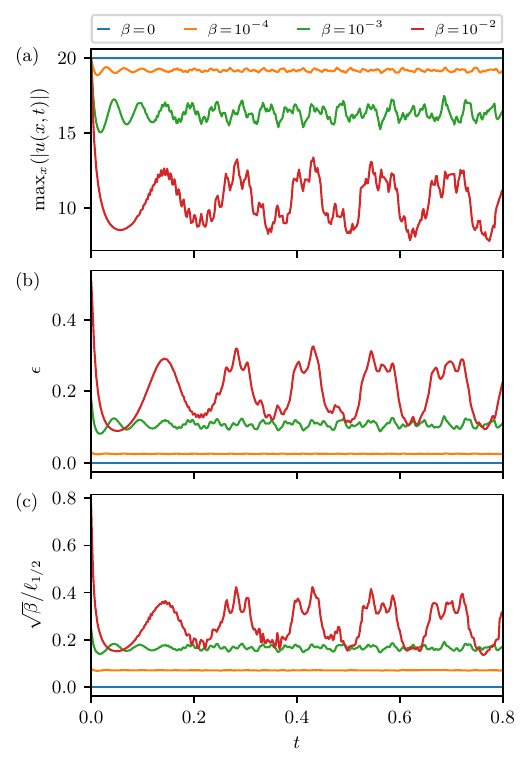}
    \caption{(a) Evolution of $\max_x(|u(x,t)|)$ for a sech-profile soliton launched into the SHE with increasing values of $\beta=0, 10^{-4}, 10^{-3}, 10^{-2}$. 
    (b) Corresponding values of the global nonintegrability measure $\epsilon = 1- H_4/H_4^{\rm NLSE}$ (defined in main text) based on the quartic contributions of the Hamiltonian. 
    (c) Local nonintegrability measure $\sqrt{\beta}/\ell_{1/2}$ (defined in main text) based on the soliton width.
    \label{fig:EvolAmplDiffG}}
\end{figure}

All the results in the main body of this paper were obtained with the nonlocality parameter $\beta=10^{-2}$. In this Appendix we examine the results of varying $\beta$ and give metrics for the deviation of the SHE away from integrability. 

We return to the
numerical experiments detailed in Sec.~\ref{sec:1soli}, and examine the evolution of a single NLSE soliton~\eqref{eq:NLSE_soli} with parameters $(A,v,s,\phi)=(20,4,\pi,0)$, launched into the SHE with four different values of $\beta= 0, 10^{-4}, 10^{-3}, 10^{-2}$. 
For all values of $\beta$ in our experiments, we observe that the injected soliton remains spatially coherent as it moves through the system. However, in the cases where $\beta\neq 0$, the deviation from the NLSE leads to the initial emission of waves and a relaxation of the soliton profile, followed by a rebound and subsequent oscillations about a new mean value, just as we reported in Sec.~\ref{sec:1soli} for $\beta=10^{-2}$, c.f.\   Fig.~\ref{fig:1soli_xt}(a). 
As we 
reduce
$\beta$ from $10^{-2}$, the initial fall in amplitude of the soliton is less dramatic, the mean value of the subsequent oscillations is closer to the 
initial
amplitude of $A=20$, the oscillation period is smaller, and its excursions are less wide. This is seen in Figure~\ref{fig:EvolAmplDiffG}(a) which shows the time evolution of the soliton peak, $\max_x(|u(x,t)|)$. 

This mollifying of the initial collapse and subsequent oscillations is natural as the deviation from integrability reduces with $\beta$. 
To quantify this, in Figure~\ref{fig:EvolAmplDiffG}(b) we plot $\epsilon = 1- H_4/H_4^{\rm NLSE}$, where $H_4^{\rm NLSE} = -(1/2)\int |u|^4 \ dx$, is the quartic energy of the NLSE, evaluated at the field configuration $u(x,t)$ of the SHE. 
The value of $\epsilon$ quantifies the global deviation of the SHE away from the NLSE, i.e., $\epsilon=0$ implies no deviation. 
Additionally, in Figure~\ref{fig:EvolAmplDiffG}(c) we show the ratio of the nonlocality lengthscale, $\sqrt{\beta}$, to $\ell_{1/2}$, the full-width half-maximum of the bound state that self-assembles in the system. This ratio is a local measure of the deviation from integrability associated with the solitonic bound state; referring to Eq.~\eqref{eq:SHE_helm_fourier}, we should have $\sqrt{\beta}/\ell_{1/2} \ll 1$ for the local nonintegrability to be small.

Figure~\ref{fig:EvolAmplDiffG} demonstrates that for $\beta < 10^{-2}$ the SHE remains weakly nonintegrable, as measured by the diagnostics $\epsilon$ and $\sqrt{\beta}/\ell_{1/2}$, for this initial condition. 
For $\beta=10^{-2}$ we consider the weakness of nonintegrability marginal; certainly for larger $\beta$ the techniques and observations we have reported on in this paper would start to enter a strongly nonintegrable regime. 
For smaller $\beta$ the dynamics associated with nonintegrability are slower (e.g.\ the two-soliton bound state takes longer to form out of soliton turbulence, and two-soliton mergers) and weaker (e.g.\ for the one-soliton initial condition reported here, for $\beta=10^{-3}$ and $10^{-4}$, $\zeta_2$ remains within the grey region denoting solitons too wide to be realised within the domain).

It is precisely these observations that motivate us to study the $\beta=10^{-2}$ case in the main body of this paper. 
Our assessment is that this value strikes a good balance between accessing novel and interesting dynamics, and retaining enough contact with the NLSE to enable us to use the DST.

\section{Physicality threshold in the DST}
\label{app:DSTnoisefloor}

In Sec.~\ref{subsubsec:DST} we noted that the Fourier collocation method we employ to 
carry out
the DST generates spurious eigenvalues close to the real axis. To highlight this, we defined a threshold $\zeta^{\rm Im}_{\rm th}$, corresponding to an NLSE soliton whose full-width half-maximum is $L/4$. Solitons with imaginary parts smaller than $\zeta^{\rm Im}_{\rm th}$ would be wider than $L/4$, and so their tails extend over our periodic domain and self-interact. We therefore discard the diagnostic information of DST eigenvalues falling beneath the threshold as being unphysical.

Other authors devise more numerically-motivated thresholds that are based on either extending $u(x,t)$ spatially by padding it with zeros, or decimating the field~\citep{agafontsev_bound-state_2023}.
Taking the DST spectrum of the new field and comparing it the original spectrum defines a new threshold, below which the eigenvalues show significant deviation. We have checked this method on representative states of the field that we report in this work, padding the field to length $3L/2$ and decimating it to $3L/4$ and taking the DST spectra. In all cases, the thresholds we calculate by comparing these spectra to the original spectrum agree with $\zeta^{\rm Im}_{\rm th}$ by an order $1$ constant. We retain our method of calculating the threshold as it is physically motivated and easy to compute, and the results of this paper depend on the location and movement of eigenvalues that are far from the threshold.

\begin{figure*}[t]
    \centering
    \includegraphics[width=0.99\linewidth]{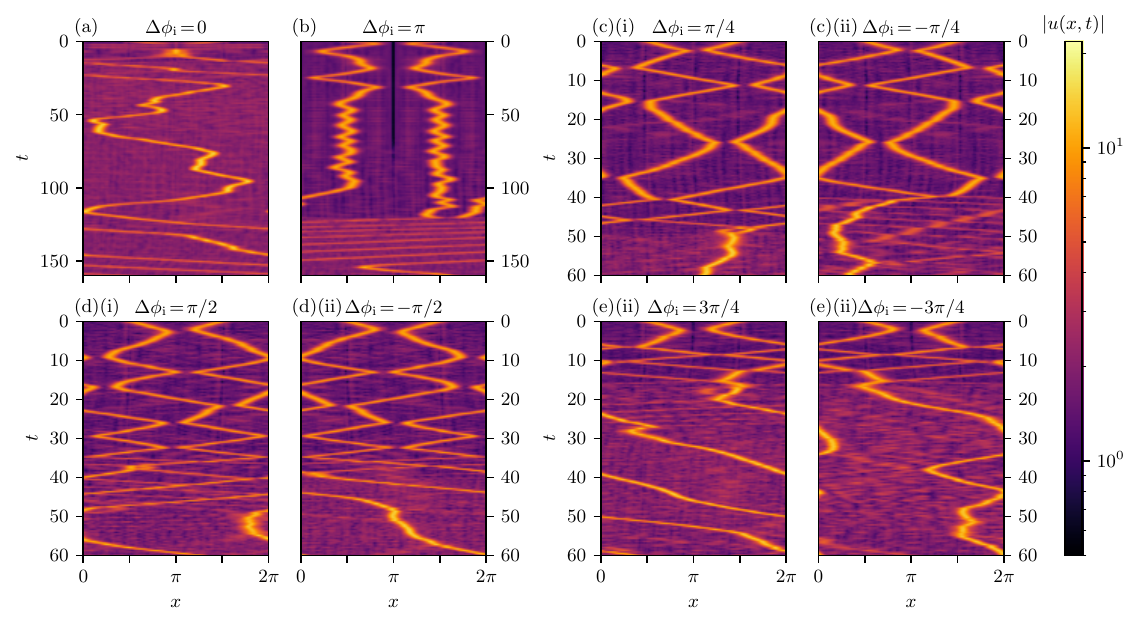}
    \caption{As per Fig.~\ref{fig:2SHE_solis_dPhi_scan}
    for two NLSE solitons launched into the SHE.}
    \label{fig:2NLS_solis_dPhi_scan}
\end{figure*}

\section{Soliton mergers---two NLSE solitons launched into the SHE}
\label{app:2NLS_solis_in_SHE} 

In Sec.~\ref{sec:2solicolls} we presented the results of two SHE solitons colliding, and showed that phase synchronisation is necessary for them to merge on collision. To come to these conclusions we have studied the collisions of many other coherent structures. 

As an example, in Fig.~\ref{fig:2NLS_solis_dPhi_scan} we present a study that is similar to Fig.~\ref{fig:2SHE_solis_dPhi_scan}, but initialised with two NLSE solitons with amplitudes $A=20$ and velocities $v=\pm 0.5$. We scan over the initial phase differences between the solitons and see once more that mergers are promoted for $\Delta\phi_{\rm i}=0$, inhibited for $\Delta\phi_{\rm i}=\pi$, and occur at intermediate times for intermediate $\Delta\phi_{\rm i}$. 
We have carried out similar analysis to Fig.~\ref{fig:2SHE_dPhi_study_1pi__4} and found once more that phase synchronisation is necessary for the merger of solitonic structures, resulting in a bound state, as was observed in Sec.~\ref{sec:2solicolls}.

Some differences between Fig.~\ref{fig:2NLS_solis_dPhi_scan} and Fig.~\ref{fig:2SHE_solis_dPhi_scan} are apparent. As the NLSE solitons initially radiate waves, as discussed in Sec.~\ref{sec:1soli}, the background is much noisier in Fig.~\ref{fig:2NLS_solis_dPhi_scan} than in Fig.~\ref{fig:2SHE_solis_dPhi_scan}. The break in reflection--phase inversion symmetry also occurs earlier, probably because the amplitudes of the injected solitons are bigger and so their collisions, whether they result in mergers or not, lead to large-amplitude spikes in the field, for which the nonlinearity is large. These lead to comparatively large numerical differences between 
nominally
symmetrical setups, which then get amplified due to chaotic dynamics.

\section*{Data availability}

No data was used for the research described in the article.

\section*{Acknowledgements}

This work was supported by the European Union's Horizon 2020 research and innovation programme under the Marie Sk\l odowska-Curie grant agreement No. 823937 for the RISE project HALT, 
and by the Simons Foundation Collaboration grant Wave Turbulence (Award ID 651471). 
	J.L. and J.S. are supported by the Leverhulme Trust Project Grant RPG-2021-014.
Some calculations were performed using the Sulis Tier 2 HPC platform hosted by the Scientific Computing Research Technology Platform at the University of Warwick. Sulis is funded by EPSRC Grant EP/T022108/1 and the HPC Midlands+ consortium.





 \bibliographystyle{elsarticle-num} 
\bibliography{bib_SHE_soli_dynamics}

\begin{thebibliography}{10}
\expandafter\ifx\csname url\endcsname\relax
  \def\url#1{\texttt{#1}}\fi
\expandafter\ifx\csname urlprefix\endcsname\relax\def\urlprefix{URL }\fi
\expandafter\ifx\csname href\endcsname\relax
  \def\href#1#2{#2} \def\path#1{#1}\fi

\bibitem{newellmoloney1992nonlinearopticsbook}
A.~C. Newell, J.~V. Moloney, Nonlinear Optics, ATIMS Series, Addison-Wesley, Redwood City, CA, 1992.

\bibitem{boyd2008nonlinearopticsbook}
R.~W. Boyd, Nonlinear Optics, 3rd Edition, Academic Press, 2008.

\bibitem{Zabusky_65}
N.~J. Zabusky, M.~D. Kruskal, Interaction of "solitons" in a collisionless plasma and the recurrence of initial states, Phys. Rev. Lett. 15 (1965) 240--243.
\newblock \href {https://doi.org/10.1103/PhysRevLett.15.240} {\path{doi:10.1103/PhysRevLett.15.240}}.

\bibitem{Bortolozzo2009optical}
U.~Bortolozzo, J.~Laurie, S.~Nazarenko, S.~Residori, Optical wave turbulence and the condensation of light, J. Opt. Soc. Am. B 26~(12) (2009).
\newblock \href {https://doi.org/10.1364/JOSAB.26.002280} {\path{doi:10.1364/JOSAB.26.002280}}.

\bibitem{Laurie_12}
J.~Laurie, U.~Bortolozzo, S.~Nazarenko, S.~Residori, One-dimensional optical wave turbulence: Experiment and theory, Phys. Rep. 514~(4) (2012) 121--175.
\newblock \href {https://doi.org/https://doi.org/10.1016/j.physrep.2012.01.004} {\path{doi:https://doi.org/10.1016/j.physrep.2012.01.004}}.

\bibitem{zakharov1988soliton}
V.~E. Zakharov, A.~N. Pushkarev, V.~F. Shvets, V.~V. Yan’kov, Soliton turbulence, JETP Lett. 48~(2) (1988) 79--82.

\bibitem{Jordan2000meanfield}
R.~Jordan, B.~Turkington, C.~L. Zirbel, A mean-field statistical theory for the nonlinear {S}chr{\"o}dinger equation, Phys. D: Nonlinear Phenom. 137~(3-4) (2000) 353--378.
\newblock \href {https://doi.org/10.1016/S0167-2789(99)00194-3} {\path{doi:10.1016/S0167-2789(99)00194-3}}.

\bibitem{Rumpf2001coherent}
B.~Rumpf, A.~C. Newell, Coherent structures and entropy in constrained, modulationally unstable, nonintegrable systems, Phys. Rev. Lett. 87 (2001) 054102.
\newblock \href {https://doi.org/10.1103/PhysRevLett.87.054102} {\path{doi:10.1103/PhysRevLett.87.054102}}.

\bibitem{castillo1996formation}
M.~D.~I. Castillo, J.~J. S{\'a}nchez-Mondrag{\'o}n, S.~Stepanov, Formation of steady-state cylindrical thermal lenses in dark stripes, Opt. Lett. 21~(20) (1996) 1622--1624.
\newblock \href {https://doi.org/10.1364/OL.21.001622} {\path{doi:10.1364/OL.21.001622}}.

\bibitem{Bekenstein2015_OpticalNewtSchro}
R.~Bekenstein, R.~Schley, M.~Mutzafi, C.~Rotschild, M.~Segev, Optical simulations of gravitational effects in the {N}ewton-{S}chr{\"o}dinger system, Nat. Phys. 11 (2015) 872--878.
\newblock \href {https://doi.org/10.1038/nphys3451} {\path{doi:10.1038/nphys3451}}.

\bibitem{Faccio2016_OpticalNewtSchro}
T.~Roger, C.~Maitland, K.~Wilson, N.~Westerberg, D.~Vocke, E.~M. Wright, D.~Faccio, Optical analogues of the {N}ewton-{S}chr{\"o}dinger equation and boson star evolution, Nat. Commun. 7 (2016) 13492.
\newblock \href {https://doi.org/10.1038/ncomms13492} {\path{doi:10.1038/ncomms13492}}.

\bibitem{conti2003route}
C.~Conti, M.~Peccianti, G.~Assanto, Route to nonlocality and observation of accessible solitons, Phys. Rev. Lett. 91~(7) (2003) 073901.
\newblock \href {https://doi.org/10.1103/PhysRevLett.91.073901} {\path{doi:10.1103/PhysRevLett.91.073901}}.

\bibitem{peccianti2003optical}
M.~Peccianti, C.~Conti, G.~Assanto, Optical modulational instability in a nonlocal medium, Phys. Rev. E 68~(2) (2003) 025602.
\newblock \href {https://doi.org/10.1103/PhysRevE.68.025602} {\path{doi:10.1103/PhysRevE.68.025602}}.

\bibitem{drazin1989solitonsbook}
P.~G. Drazin, R.~S. Johnson, Solitons: an introduction, Vol.~2 of Cambridge texts in applied mathematics, Cambridge University Press, 1989.
\newblock \href {https://doi.org/10.1017/CBO9781139172059} {\path{doi:10.1017/CBO9781139172059}}.

\bibitem{jia2012solitons}
J.~Jia, J.~Lin, Solitons in nonlocal nonlinear kerr media with exponential response function, Opt. Express 20~(7) (2012) 7469--7479.
\newblock \href {https://doi.org/10.1364/OE.20.007469} {\path{doi:10.1364/OE.20.007469}}.

\bibitem{horikis2020exact}
T.~P. Horikis, Exact solutions and self-similar symmetries of a nonlocal nonlinear {S}chr{\"o}dinger equation, Eur. Phys. J. Plus 135~(7) (2020) 562.
\newblock \href {https://doi.org/10.1140/epjp/s13360-020-00571-w} {\path{doi:10.1140/epjp/s13360-020-00571-w}}.

\bibitem{zakharov2004_1DWT}
V.~Zakharov, F.~Dias, A.~Pushkarev, One-dimensional wave turbulence, Phys. Rep. 398~(1) (2004) 1--65.
\newblock \href {https://doi.org/10.1016/j.physrep.2004.04.002} {\path{doi:10.1016/j.physrep.2004.04.002}}.

\bibitem{Zakharov_72}
V.~E. Zakharov, A.~B. Shabat, Exact theory of two-dimensional self-focusing and one-dimensional self-modulation of waves in nonlinear media, J. Exp. Theor. Phys. 34~(1) (1972) 62--69.

\bibitem{Ablowitz_book}
M.~J. Ablowitz, H.~Segur, Solitons and the Inverse Scattering Transform, Society for Industrial and Applied Mathematics, 1981.
\newblock \href {https://doi.org/10.1137/1.9781611970883} {\path{doi:10.1137/1.9781611970883}}.

\bibitem{gardner1967method}
C.~S. Gardner, J.~M. Greene, M.~D. Kruskal, R.~M. Miura, Method for solving the {K}orteweg-de{V}ries equation, Phys. Rev. Lett. 19~(19) (1967) 1095.
\newblock \href {https://doi.org/10.1103/PhysRevLett.19.1095} {\path{doi:10.1103/PhysRevLett.19.1095}}.

\bibitem{Turitsyn_17}
S.~K. Turitsyn, J.~E. Prilepsky, S.~T. Le, S.~Wahls, L.~L. Frumin, M.~Kamalian, S.~A. Derevyanko, Nonlinear {F}ourier transform for optical data processing and transmission: advances and perspectives, Optica 4~(3) (2017) 307--322.
\newblock \href {https://doi.org/10.1364/OPTICA.4.000307} {\path{doi:10.1364/OPTICA.4.000307}}.

\bibitem{sugavanam_analysis_2019}
S.~Sugavanam, M.~K. Kopae, J.~Peng, J.~E. Prilepsky, S.~K. Turitsyn, Analysis of laser radiation using the {Nonlinear} {Fourier} transform, Nat. Commun. 10~(1) (2019) 5663.
\newblock \href {https://doi.org/10.1038/s41467-019-13265-4} {\path{doi:10.1038/s41467-019-13265-4}}.

\bibitem{el2021soliton}
G.~A. El, Soliton gas in integrable dispersive hydrodynamics, J. Stat.Mech. Theory Exp. 2021~(11) (2021) 114001.
\newblock \href {https://doi.org/10.1088/1742-5468/ac0f6d} {\path{doi:10.1088/1742-5468/ac0f6d}}.

\bibitem{suret2024soliton}
P.~Suret, S.~Randoux, A.~Gelash, D.~Agafontsev, B.~Doyon, G.~El, Soliton gas: Theory, numerics, and experiments, Phys. Rev. E 109~(6) (2024) 061001.
\newblock \href {https://doi.org/10.1103/PhysRevE.109.061001} {\path{doi:10.1103/PhysRevE.109.061001}}.

\bibitem{orszag1969numerical}
S.~A. Orszag, Numerical methods for the simulation of turbulence, Phys. Fluids 12~(12) (1969) II--250.
\newblock \href {https://doi.org/10.1063/1.1692445} {\path{doi:10.1063/1.1692445}}.

\bibitem{gottlieb1977numericalBOOK}
D.~Gottlieb, S.~A. Orszag, Numerical analysis of spectral methods: theory and applications, SIAM, 1977.
\newblock \href {https://doi.org/10.1137/1.9781611970425} {\path{doi:10.1137/1.9781611970425}}.

\bibitem{krstulovic2011dispersive}
G.~Krstulovic, M.~Brachet, Dispersive bottleneck delaying thermalization of turbulent {Bose}-{Einstein} condensates, Phys. Rev. Lett. 106~(11) (2011) 115303.
\newblock \href {https://doi.org/10.1103/PhysRevLett.106.115303} {\path{doi:10.1103/PhysRevLett.106.115303}}.

\bibitem{cox2002exponential}
S.~M. Cox, P.~C. Matthews, Exponential time differencing for stiff systems, J. Comput. Phys. 176~(2) (2002) 430--455.
\newblock \href {https://doi.org/10.1006/jcph.2002.6995} {\path{doi:10.1006/jcph.2002.6995}}.

\bibitem{Boffetta_92}
G.~Boffetta, A.~R. Osborne, Computation of the direct scattering transform for the nonlinear {S}chroedinger equation, J. Comput. Phys. 102~(2) (1992) 252--264.
\newblock \href {https://doi.org/https://doi.org/10.1016/0021-9991(92)90370-E} {\path{doi:https://doi.org/10.1016/0021-9991(92)90370-E}}.

\bibitem{Yang_book_2010}
J.~Yang, Nonlinear Waves in Integrable and Nonintegrable Systems, Society for Industrial and Applied Mathematics, 2010.
\newblock \href {https://doi.org/10.1137/1.9780898719680} {\path{doi:10.1137/1.9780898719680}}.

\bibitem{agafontsev_bound-state_2023}
D.~S. Agafontsev, A.~A. Gelash, R.~I. Mullyadzhanov, V.~E. Zakharov, Bound-state soliton gas as a limit of adiabatically growing integrable turbulence, Chaos Solitons Fractals 166 (2023) 112951.
\newblock \href {https://doi.org/10.1016/j.chaos.2022.112951} {\path{doi:10.1016/j.chaos.2022.112951}}.

\bibitem{nazarenko2006wave}
S.~Nazarenko, M.~Onorato, Wave turbulence and vortices in {B}ose-{E}instein condensation, Phys. D: Nonlinear Phenom. 219~(1) (2006) 1--12.
\newblock \href {https://doi.org/10.1016/j.physd.2006.05.007} {\path{doi:10.1016/j.physd.2006.05.007}}.

\bibitem{diLeoni2015spatio}
P.~C. di~Leoni, P.~J. Cobelli, P.~D. Mininni, The spatio-temporal spectrum of turbulent flows, Eur. Phys. J. E 38~(12) (2015) 1--10.
\newblock \href {https://doi.org/10.1140/epje/i2015-15136-x} {\path{doi:10.1140/epje/i2015-15136-x}}.

\bibitem{osborne1998soliton}
A.~R. Osborne, M.~Onorato, M.~Serio, L.~Bergamasco, Soliton creation and destruction, resonant interactions, and inelastic collisions in shallow water waves, Phys. Rev. Lett. 81~(17) (1998) 3559.
\newblock \href {https://doi.org/10.1103/PhysRevLett.81.3559} {\path{doi:10.1103/PhysRevLett.81.3559}}.

\bibitem{kuznetsov_nonlinear_1995}
E.~A. Kuznetsov, A.~V. Mikhailov, I.~A. Shimokhin, Nonlinear interaction of solitons and radiation, Phys. D: Nonlinear Phenom. 87~(1) (1995) 201--215.
\newblock \href {https://doi.org/10.1016/0167-2789(95)00149-X} {\path{doi:10.1016/0167-2789(95)00149-X}}.

\bibitem{kachulin2021bound}
D.~Kachulin, S.~Dremov, A.~Dyachenko, Bound coherent structures propagating on the free surface of deep water, Fluids 6~(3) (2021) 115.
\newblock \href {https://doi.org/https://doi.org/10.3390/fluids6030115} {\path{doi:https://doi.org/10.3390/fluids6030115}}.

\bibitem{gelash2024bisolitons}
A.~Gelash, S.~Dremov, R.~Mullyadzhanov, D.~Kachulin, Bi-solitons on the surface of a deep fluid: An inverse scattering transform perspective based on perturbation theory, Phys. Rev. Lett. 132~(13) (2024) 133403.
\newblock \href {https://doi.org/https://doi.org/10.1103/PhysRevLett.132.133403} {\path{doi:https://doi.org/10.1103/PhysRevLett.132.133403}}.

\bibitem{agafontsev2024multisoliton}
D.~Agafontsev, A.~Gelash, S.~Randoux, P.~Suret, Multisoliton interactions approximating the dynamics of breather solutions, Stud. Appl. Math. 152~(2) (2024) 810--834.
\newblock \href {https://doi.org/10.1111/sapm.12662} {\path{doi:10.1111/sapm.12662}}.

\bibitem{soto-crespo_integrable_2016}
J.~M. Soto-Crespo, N.~Devine, N.~Akhmediev, Integrable {Turbulence} and {Rogue} {Waves}: {Breathers} or {Solitons}?, Phys. Rev. Lett. 116~(10) (2016) 103901.
\newblock \href {https://doi.org/10.1103/PhysRevLett.116.103901} {\path{doi:10.1103/PhysRevLett.116.103901}}.

\bibitem{kaup1976perturbation}
D.~J. Kaup, A perturbation expansion for the {Z}akharov-{S}habat inverse scattering transform, SIAM J. Appl. Math. 31~(1) (1976) 121--133.
\newblock \href {https://doi.org/https://doi.org/10.1137/0131013} {\path{doi:https://doi.org/10.1137/0131013}}.

\bibitem{karpman1977perturbation}
V.~I. Karpman, E.~M. Maslov, Perturbation theory for solitons, Zh. Eksp. Teor. Fiz. 73~(2) (1977) 537--559.

\bibitem{kivshar1989dynamics}
Y.~S. Kivshar, B.~A. Malomed, Dynamics of solitons in nearly integrable systems, Rev. Modern Phys. 61~(4) (1989) 763.
\newblock \href {https://doi.org/https://doi.org/10.1103/RevModPhys.61.763} {\path{doi:https://doi.org/10.1103/RevModPhys.61.763}}.

\bibitem{nazarenko2011book}
S.~Nazarenko, Wave {Turbulence}, no. 825 in Lecture {Notes} in {Physics}, Springer-Verlag, Berlin, Heidelberg, 2011.
\newblock \href {https://doi.org/10.1007/978-3-642-15942-8} {\path{doi:10.1007/978-3-642-15942-8}}.

\end{thebibliography}





\end{document}